\documentclass[12pt]{iopart}

\usepackage{psfig}

\def\beq{\begin{equation}}
\def\eeq{\end{equation}}
\def\ber{\begin{eqnarray}}
\def\eer{\end{eqnarray}}
\def\m{{\rm m}}
\def\C{{\cal C}}

\def\bc{A}

\def\lsim{\
  \lower-2.0pt\vbox{\hbox{\rlap{$<$}\lower5.5pt\vbox{\hbox{$\sim$}}}}\ }
\def\gsim{\
  \lower-2.0pt\vbox{\hbox{\rlap{$>$}\lower5.5pt\vbox{\hbox{$\sim$}}}}\ }

\begin{document}

\title{Gravitational instability on the brane: the role of boundary conditions}

\author{Yuri Shtanov$^a$, Alexander Viznyuk$^a$ and Varun Sahni$^b$}

\address{$^a$Bogolyubov Institute for Theoretical Physics, Kiev 03680,
Ukraine}
\address{$^b$Inter-University Centre for Astronomy and Astrophysics,
Post Bag 4, Ganeshkhind, Pune 411~007, India}
\eads{\mailto{shtanov@bitp.kiev.ua}, \mailto{viznyuk@bitp.kiev.ua},
\mailto{varun@iucaa.ernet.in}}

\begin{abstract}
\noindent An outstanding issue in braneworld theory concerns the setting up of
proper boundary conditions for the brane--bulk system. Boundary conditions
(BC's) employing regulatory branes or demanding that the bulk metric be
nonsingular have yet to be implemented in full generality.  In this paper, we
take a different route and specify boundary conditions directly on the brane
thereby arriving at a local and closed system of equations (on the brane). We
consider a one-parameter family of boundary conditions involving the
anisotropic stress of the projection of the bulk Weyl tensor on the brane and
derive an exact system of equations describing scalar cosmological
perturbations on a generic braneworld with induced gravity. Depending upon our
choice of boundary conditions, perturbations on the brane either grow
moderately (region of stability) or rapidly (instability). In the instability
region, the evolution of perturbations usually depends upon the scale: small
scale perturbations grow much more rapidly than those on larger scales. This
instability is caused by a peculiar gravitational interaction between dark
radiation and matter on the brane. Generalizing the boundary conditions
obtained by Koyama and Maartens, we find for the Dvali--Gabadadze--Porrati
model an instability, which leads to a dramatic scale-dependence of the
evolution of density perturbations in matter and dark radiation. A different
set of BC's, however, leads to a more moderate and scale-independent growth of
perturbations. For the {\em mimicry\/} braneworld, which expands like LCDM,
this class of BC's can lead to an earlier epoch of structure formation.
\end{abstract}


\maketitle

\section{Introduction}

\begin{figure}
\begin{center}
\centerline{\psfig{figure=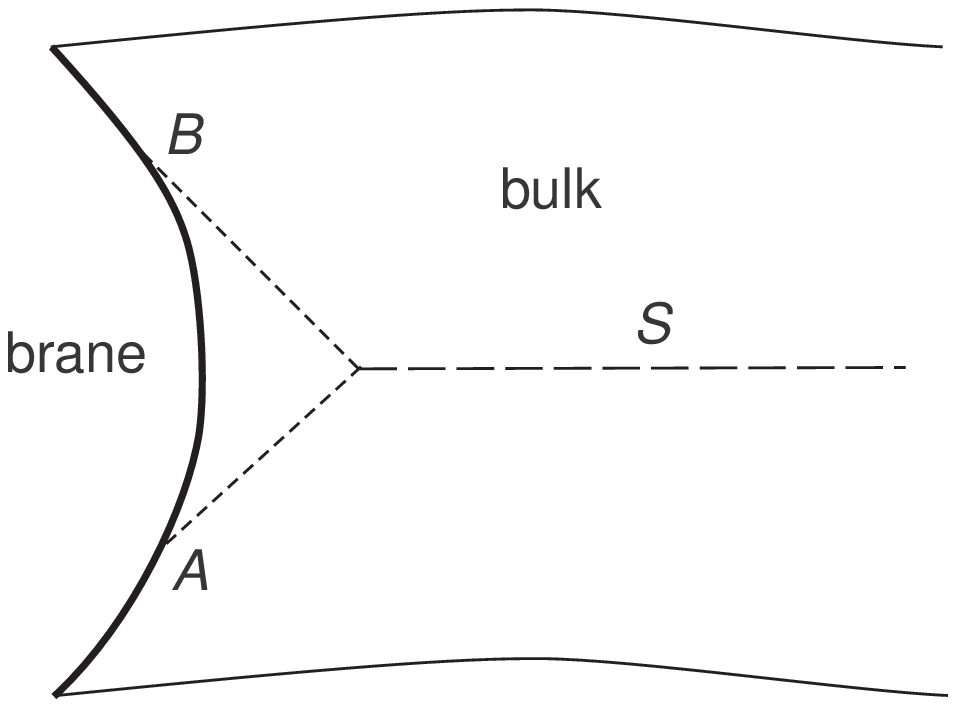,width=0.55\textwidth,angle=0} }
\end{center}
\caption{\small {\bf Brane as a four-dimensional boundary of the
five-dimensional bulk.} \newline  By smoothly perturbing the data on a
hypersurface $S$ in the bulk at some distance from the brane, one modifies the
solution on the brane only to the past of the hypersurface $A$ and to the
future of the hypersurface $B$; hence, even complete knowledge of the data on
the brane in the spacetime region between the hypersurfaces $A$ and $B$ is not
sufficient for predicting the evolution to the past or future of this
region.\label{fig:brane}}
\end{figure}

A specific feature of braneworld models of gravity \cite{RS,DGP} and cosmology
\cite{BDL,cosmol,DDG} is that the arena of observable events occurs in a
lower-dimensional Lorentzian submanifold of the higher-dimensional space-time.
For instance, in the scenario which we shall be considering, the brane is a
four-dimensional boundary of a space-time with a noncompact (`infinite') extra
dimension, and an observer is in direct contact only with the induced metric on
the brane.  In this situation, one might worry about the well-posedness of
physical problems from the observer's viewpoint. Indeed, phenomena taking place
on the brane can depend upon physical conditions in the higher-dimensional bulk
manifold which is inaccessible to an observer living on the brane.
Consequently, the brane observer loses the ability of predicting classical
physical phenomena on the brane. Indeed, it appears that no initial-value
problem can be well posed on the brane since, no matter how detailed the
specification of the initial conditions on a spacelike hypersurface on the
brane, this does not uniquely determine events occurring either in the future
or the past of this hypersurface. At best, one can uniquely predict the events
in the future of a Cauchy hypersurface on the brane if one has the knowledge of
the {\em entire past} to this hypersurface (and vice versa). This situation is
illustrated by Fig.~\ref{fig:brane}, which depicts the brane as a boundary of a
bulk space with a noncompact extra dimension. Let a solution of the bulk--brane
system be given. By smoothly perturbing data on a hypersurface $S$ in the bulk
at some distance from the brane, one modifies the solution on the brane only to
the past of the hypersurface $A$ and to the future of the hypersurface $B$;
hence, even complete knowledge of the data on the brane in the region between
the hypersurfaces $A$ and $B$ is not sufficient for predicting the evolution to
the past or future of this region.  This is sometimes expressed by saying that
the braneworld equations are {\em nonlocal\/} from the viewpoint of an observer
living on the brane.\footnote[1]{Similar effects can also occur in our
four-dimensional world. For an observer bound to the Earth, any local
experiment has an uncertainty in its outcome because the events which lie
outside the observer's past light cone are totally unknown but, in principle,
can intervene the experiment at any moment of time. Thus, an unwanted cosmic
gamma ray or gravity wave can suddenly affect the experimental situation. One
can shield the experimental device from gamma rays, but not from gravity waves.
Perhaps, we are fortunate that gravity waves (if they exist) are so feeble that
even to detect them constitutes a big problem.  In the opposite extreme case,
were we permanently subject to gravity waves of strong amplitudes propagating
in all directions, all our science of celestial motion would probably be
useless (if not unattainable), just as the science of marine navigation becomes
useless during a big storm.}

One might try to limit the space of solutions by specifying some appropriate
boundary conditions for the full brane--bulk system. Much effort has been
applied to formulate reasonable conditions in the bulk, by demanding that the
bulk metric be nonsingular or by employing other (regulatory) branes. However,
to date, these proposals have not been implemented in the braneworld theory in
full generality --- with certain success they were used only in a linearized
theory under some additional approximations (see e.g., \cite{LSS,KM,SSH}) ---
and it is clear that such an approach to boundary conditions still leaves
intact the property of nonlocality described above.

The issue of cosmological perturbations on the brane in induced-gravity models
remains poorly studied in the literature precisely because of the difficulties
posed by the nonlocal character of this theory and the lack of proper boundary
conditions. Thus, in treating the scalar perturbations on the brane, one is led
to consider the problem in the bulk space using, for instance, the Mukohyama
master variable \cite{Mukohyama}, which is a scalar function describing
gravitational degrees of freedom in the bulk and obeying there a second-order
partial differential equation.  The corresponding initial-value problem was
argued to be well posed in the brane--bulk system \cite{Deffayet}. However, in
this case, one needs to specify initial conditions on the brane as well as in
the bulk, with the obvious complication of having to deal with a function
dependent on the coordinate of the extra dimension. This problem can be tackled
to some extent in the linear approximation, but looks insurmountable when
nonlinearities in the metric become significant --- and, in principle, they can
become significant in the bulk space much `earlier' than on the brane.  And
--- what is more important --- one still has to face the problem of boundary
conditions in the bulk, since not all possible values of the Mukohyama variable
may lead to well-behaved solutions of the brane-bulk system.

In an earlier paper \cite{SSV}, we suggested a new approach to the issue of
boundary conditions for the brane--bulk system.  From a broader perspective,
boundary conditions can be regarded as any conditions which restrict the space
of solutions. Our idea was to specify such conditions directly on the brane
which represents the observable world, so as to arrive at a local and closed
system of equations on the brane. In fact, this is what some researches usually
do in practice in the form of making various reasonable assumptions (see, e.g.,
\cite{Maartens,BM} for the issue of cosmological perturbations; and \cite{bh}
for the case of spherically symmetric solutions on the brane). The behaviour of
the metric in the bulk is of no further concern in this approach, since this
metric is, for all practical purposes, unobservable directly. As first noted in
\cite{SMS}, the nonlocality of the braneworld equations is connected with the
dynamical properties of the bulk Weyl tensor projected on to the brane. It
therefore seems logical to impose certain restrictions on this tensor in order
to obtain a closed system of equations on the brane.\footnote{We admit that our
current approach is quite radical because it effectively ``freezes'' certain
degrees of freedom in the bulk; but its merit is that it apparently leads to a
well-defined closed, local, causal, and, in principle, verifiable theory of
gravity in four dimensions. If this proposal of specifying boundary conditions
on the brane seems too radical, then one can view the results of this paper as
an investigation of a certain class of approximations to the perturbation
equations on the brane, some of which have appeared in the literature
\cite{KM,SSH}.}

Neglecting the projected Weyl tensor on the brane \cite{wrong}, though simple,
is incorrect \cite{KM} since it is incompatible with the equation that follows
from the Bianchi identity [see Eq.~(\ref{conserv}) below]. Hence, to proceed
along this road, one needs to impose some other condition on this tensor.
Perhaps, the simplest choice is to set to zero its (appropriately defined)
anisotropic stress. This condition is fully compatible with all equations of
the theory and eventually amounts to a brane universe described by a modified
theory of gravity with an additional invisible component --- the Weyl fluid, or
dark radiation --- having nontrivial dynamics. However, this is not a unique
prescription and we shall discuss other possibilities in this paper applying
some of them to the study of scalar cosmological perturbations on the brane.

An important result of this paper is that our choice of boundary conditions
(BC's) strongly influences the growth of density perturbations. For a class of
BC's, a qualitative analysis of the linearized equations for scalar
perturbations in a matter-dominated braneworld reveals the existence of an
instability. In particular, the boundary condition which formally coincides
with the approximate condition derived in \cite{KM} in the context of the
Dvali--Gabadadze--Porrati (DGP) model \cite{DGP}, leads to a dramatic
dependence of the growth of perturbations on spatial scale. In a
matter-dominated braneworld, perturbations on small spatial scales grow many
orders of magnitude larger than those on large spatial scales, indicating an
early breakdown of linear regime. The cause of this instability is a peculiar
gravitational interaction between matter and dark radiation on the brane which
arises for this class of BC's. Our qualitative conclusions are supported by a
numerical integration of the exact system of linearized equations.
A different choice of BC's, however, leads to a perfectly `normal' situation in
which the growth of perturbations is moderate and virtually independent of
scale.

In the DGP model, the brane tension and the bulk cosmological constant are both
set to zero, and the current acceleration of the universe expansion is
explained as an effect of extra dimension without severe fine tuning of the
fundamental constants of the theory.  However, an interesting situation arises
also in the opposite case, where both these constants are large. As shown in
\cite{SSV}, a low-density braneworld exactly mimics the expansion properties of
the LCDM model. In particular, a universe consisting solely of baryons with
$\Omega_{\rm b} \simeq 0.04$ can mimic the LCDM cosmology with a much larger
`effective' value of the matter density $\Omega_\m \simeq 0.2$$-$$0.3$. A
conventional low-density universe runs into trouble with observations on
account of its slow growth rate of perturbations. It is therefore interesting
that such a model survives in the braneworld context which allows perturbations
to grow much faster. 

Whether one can altogether do away with the notion of dark matter, as suggested
in \cite{HMC,PBK} is a moot point, since, in addition to alleviating the
so-called {\em growth problem\/}, dark matter also explains other observations
including rotation curves of galaxies, etc. Whether all such observations can
be accommodated in induced-gravity models is an interesting open subject which
we set aside for future investigations. For recent results in frames of the the
induced-gravity models, see our paper \cite{VS}.

This paper is organized as follows:  In section \ref{sec:bc}, we introduce the
theory and formulate the boundary conditions.  In section \ref{sec:perturb}, we
apply the theory to the issue of scalar cosmological perturbations. We derive
an exact convenient system of equation governing the growth of perturbations of
pressureless matter and dark radiation and discuss boundary conditions of
various type. In section \ref{sec:DGP}, we study numerically the evolution of
perturbations in the DGP cosmology, and in section \ref{sec:mimicry} we do this
for the mimicry braneworld models.  Finally, we discuss our results in section
\ref{sec:discuss}.

\section{Boundary conditions for the brane--bulk system} \label{sec:bc}

We consider a generic braneworld model with the action given by the expression
\begin{equation} \label{action}
\fl M^3 \left[\int_{\rm bulk} \left( {\cal R} - 2 \Lambda \right) - 2 \int_{\rm
brane} K \right] + \int_{\rm brane} \left( m^2 R - 2 \sigma \right) + \int_{\rm
brane} L \left( h_{ab}, \phi \right) \, .
\end{equation}
Here, ${\cal R}$ is the scalar curvature of the metric $g_{ab}$ in the
five-dimensional bulk, and $R$ is the scalar curvature of the induced metric
$h_{ab} = g_{ab} - n_a n_b$ on the brane, where $n^a$ is the vector field of
the inner unit normal to the brane, which is assumed to be a boundary of the
bulk space, and the notation and conventions of \cite{Wald} are used. The
quantity $K = h^{ab} K_{ab}$ is the trace of the symmetric tensor of extrinsic
curvature $K_{ab} = h^c{}_a \nabla_c n_b$ of the brane. The symbol $L (h_{ab},
\phi)$ denotes the Lagrangian density of the four-dimensional matter fields
$\phi$ whose dynamics is restricted to the brane so that they interact only
with the induced metric $h_{ab}$. All integrations over the bulk and brane are
taken with the corresponding natural volume elements. The symbols $M$ and $m$
denote the five-dimensional and four-dimensional Planck masses, respectively,
$\Lambda$ is the bulk cosmological constant, and $\sigma$ is the brane tension.
Note that the Hilbert--Einstein term $\int_{\rm brane} m^2 R$ may be induced by
quantum fluctuations of matter fields residing on the brane \cite{DGP,cosmol}.
This idea was originally suggested by Sakharov in the context of his {\em
induced gravity\/} model of general relativity \cite{sakharov}.

The following famous cosmologies are related to important subclasses of action
(\ref{action}):
\begin{enumerate}

\item The Randall--Sundrum model \cite{RS} is obtained after setting $m=0$ in
(\ref{action}).

\item The Dvali--Gabadadze--Porrati model \cite{DGP} corresponds to the special
case where both the cosmological constant in the bulk and the brane tension
vanish, i.e., $\Lambda = 0$ and $\sigma = 0$ in (\ref{action}).

\item Finally, general relativity, leading to the LCDM cosmological model, is
obtained after setting $M = 0$ in (\ref{action}).
\end{enumerate}

In addition to the above, braneworld cosmology described by (\ref{action}) is
very rich in possibilities and leads to several new and interesting
cosmological scenarios including the following
\cite{SSV,SS02,loiter,sahni_review}:
\begin{itemize}

\item Braneworld universe can accelerate at late times, and the effective
equation of state of dark energy can be $w_{\rm eff} \leq -1$ as well as
$w_{\rm eff} \geq -1$. The former provides an example of {\em phantom
cosmology\/} without the latter's afflictions.

\item The current acceleration of the universe is a {\em transient\/} feature
in a class of braneworld models. In such models, both the past and the future
are described by matter-dominated expansion.

\item The braneworld scenario is flexible enough to allow a spatially flat
universe to {\em loiter\/} in the past ($z \geq \mbox{few}$). During loitering,
the universe expands more slowly than in LCDM (i.e., $H(z) / H_{\rm LCDM} <
1$), which leads to interesting observational possibilities for this scenario.

\item For large values of some of its parameters, the braneworld can {\em
mimic\/} LCDM at low redshifts, so that $H(z) \simeq H_{\rm LCDM}$ and $w_{\rm
eff} \simeq -1$ for $z \leq \mbox{few}$. At higher redshifts, the Hubble
parameter departs from its value in concordance cosmology leading to important
observational possibilities.
\end{itemize}

It is therefore quite clear that the action (\ref{action}) can give rise to
cosmological models with quite definite attributes and properties. Confronting
these models against observations becomes a challenging and meaningful
exercise. At present, most tests of these models have been conducted under the
assumption that the three-dimensional brane is homogeneous and isotropic
\cite{alam06}. Since braneworld cosmology has passed these tests successfully,
the next step clearly is to probe it deeper by examining its perturbations.
This shall form the focus of the present paper.

The action (\ref{action}) leads to the Einstein equation with cosmological constant
in the bulk:
\begin{equation} \label{bulk}
{\cal G}_{ab} + \Lambda g_{ab} = 0 \, ,
\end{equation}
while the field equation on the brane is
\begin{equation} \label{brane}
m^2 G_{ab} + \sigma h_{ab} = T_{ab} + M^3 \left(K_{ab} - h_{ab} K \right) \, ,
\end{equation}
where $T_{ab}$ is the stress--energy tensor on the brane stemming from the last
term in action (\ref{action}). By using the Gauss--Codazzi identities and
projecting the field equations onto the brane, one obtains the effective equation
\cite{SSV,SMS}
\begin{equation}\label{effective}
G_{ab} + \Lambda_{\rm eff} h_{ab} = 8 \pi G_{\rm eff} T_{ab} + {1 \over \beta +
1} \left( {1 \over M^6} Q_{ab} - C_{ab} \right) \, ,
\end{equation}
where
\begin{equation}\label{beta}
\beta = {2 \sigma m^2 \over 3 M^6}
\end{equation}
is a dimensionless parameter,
\begin{equation} \label{lambda-eff}
\Lambda_{\rm eff} = {1 \over \beta + 1 } \left({\Lambda \over 2} + {\sigma^2
\over 3 M^6}\right)
\end{equation}
is the effective cosmological constant,
\begin{equation}\label{g-eff}
8 \pi G_{\rm eff} = {\beta \over \beta + 1} \cdot {1 \over m^2}
\end{equation}
is the effective gravitational constant,
\begin{equation}\label{q}
Q_{ab} = \frac13 E E_{ab} - E_{ac} E^{c}{}_b + \frac12 \left(E_{cd} E^{cd} -
\frac13 E^2 \right) h_{ab}
\end{equation}
is a quadratic expression with respect to the `bare' Einstein equation $E_{ab}
\equiv m^2 G_{ab} - T_{ab}$ on the brane, and $E = h^{ab} E_{ab}$. The
symmetric traceless tensor $C_{ab} \equiv n^c n^d C_{acbd}$ in
(\ref{effective}) is a projection of the bulk Weyl tensor $C_{abcd}$.  It is
connected with the tensor $Q_{ab}$ through the equation following from the
energy conservation and Bianchi identity:
\begin{equation}\label{conserv}
\widetilde \nabla^a \left( Q_{ab} - M^6 C_{ab} \right) = 0 \, ,
\end{equation}
where $\widetilde \nabla^a$ denotes the covariant derivative on the brane
associated with the induced metric $h_{ab}$.

The system of equations (\ref{bulk}), (\ref{brane}) can have many formal
solutions --- too many to represent physically admissible cases.  The issue
that arises is what conditions should determine the space of its physical
solutions.  A common approach consists in demanding that the bulk space be
singularity free.  This approach is physically most appealing; however, because
of obvious difficulties, it is very difficult to implement this idea in
practice.  (In the Appendix, we critically discuss the recent efforts
\cite{KM,SSH} to derive the proper boundary condition in this way.) Moreover,
it may not be really necessary to impose this condition given that the entire
observable world is represented by the brane only. Provided that the physical
situation on the brane is regular, it may not be vital to worry about the
situation in the bulk. This last consideration suggests that one might try to
impose boundary conditions for the brane--bulk system not far away in the bulk
but closer to home
--- directly on the brane or in a close neighbourhood of the brane  ---
since this approach is likely to be the one most relevant for an observer
residing in our (3+1)-dimensional universe.

It is clear that the absence of the time derivatives of certain components of
the traceless symmetric tensor $C_{ab}$ in Eq.~(\ref{conserv}) results in a
functional arbitrariness in the dynamics of this tensor on the brane. Hence, it
is reasonable to postulate boundary conditions in the form of certain
constraints imposed on this tensor.  In fact, conditions of this sort on the
components of the tensor $C_{ab}$ have been applied in many papers as an
approximation or a reasonable guess (see, e.g., \cite{Maartens,BM} for a
discussion in the Randall--Sundrum model). Here, we would like to consider such
conditions as an exact physical principle.

The tensor $C_{ab}$ can be decomposed through a convenient physically
determined normalized timelike vector field $u^a$ on the brane (see
\cite{Maartens}):
\begin{equation} \label{weyl}
m^2 C_{ab} = \rho_\C \left( u_a u_b + \frac13 q_{ab} \right) + 2 v_{(a} u_{b)}
+ \pi_{ab}\, ,
\end{equation}
where
\begin{equation}
q_{ab} = h_{ab} + u_a u_b
\end{equation}
is the tensor of projection to the tangent subspace orthogonal to $u^a$, and
the covector $v_a$ and traceless symmetric tensor $\pi_{ab}$ are both
orthogonal to $u^a$.  In this decomposition, the quantities $\rho_\C$, $v_a$,
and $\pi_{ab}$ have the meaning of dark-radiation density, momentum transfer,
and anisotropic stress, respectively, as measured by observers following the
world lines tangent to $u^a$.

The anisotropic stress can be further conveniently decomposed into scalar,
vector, and tensor parts in the following way:
\begin{equation} \label{decompose}
\pi_{ab} = \left[ D_{(a} D_{b)} - \frac13 q_{ab} D^2 \right] \pi_\C + D_{(a}
\pi_{b)} + \widetilde \pi_{ab} \, .
\end{equation}
Here, we have introduced the derivative operator $D_a$ acting on tensor fields
$T^{cd\cdots}_{ef\cdots}$ tangent to the brane and orthogonal to $u^a$
according to the rule
\begin{equation}
D_a T^{cd\cdots}_{ef\cdots} = \prod \widetilde\nabla_a T^{cd\cdots}_{ef\cdots}
\, ,
\end{equation}
where the symbol $\prod$ denotes the projection using $q_{ab}$ with respect to
all tensorial indices. For example,
\begin{equation}
D_a \pi_b = q^c{}_a q^d{}_b \widetilde \nabla_c \pi_d \, .
\end{equation}
If the vector field $u^a$ is hypersurface orthogonal, then $q_{ab}$ represents
the induced metric in the family of hypersurfaces orthogonal to $u^a$, and
$D_a$ is the (unique) derivative operator on these hypersurfaces compatible
with this induced metric.  Furthermore, the covector field $\pi_a$ and the
symmetric tensor $\widetilde \pi_{ab}$ in (\ref{decompose}) are both orthogonal
to $u^a$ and, for their unique specification, the latter can also be assumed to
have zero trace and `spatial divergence' (i.e., to be trace-free and
`transverse'):
\begin{equation}
 h^{ab} \widetilde \pi_{ab} = 0 \, , \qquad D^a \widetilde \pi_{ab} = 0 \, .
\end{equation}
The first of these conditions implies $D^a \pi_a = 0$.

There are no evolution equations for the tensor fields $\pi_{ab}$ on the brane,
which is a manifestation of the nonlocality problem discussed in the
introduction. Therefore, boundary conditions on the brane can be specified by
imposing additional conditions on this tensor. The simplest possible choice of
such conditions would be to set it entirely to zero:
\begin{equation} \label{bc0}
\pi_{ab} = 0 \, .
\end{equation}
In this case, equation (\ref{conserv}) gives the evolution equations for the
remaining dark-radiation components $\rho_\C$ and $v_a$.  However, other
conditions on the components $\pi_\C$, $\pi_a$, and $\widetilde \pi_{ab}$ of
the anisotropic stress are possible.  Without introducing any additional
dimensional parameters into the theory, one can relate the scalar `spatial
Laplacian' $D^2 \pi_\C$ to the energy density $\rho_\C$
\begin{equation} \label{bc}
D^2 \pi_\C = \bc \rho_\C \, ,
\end{equation}
while setting $\pi_a = 0$ and $\widetilde \pi_{ab} = 0$. Here, $\bc$ is some
dimensionless covariant scalar.  In this paper, we mostly assume $\bc$ to be
just a constant so that the boundary condition (\ref{bc}) forms a one-parameter
family.

Boundary conditions of the type (\ref{bc0}) or (\ref{bc}) clearly depend on the
choice of the timelike vector field $u^a$ used in the decompositions
(\ref{weyl}) and (\ref{decompose}).  In order for these BC's to be covariantly
specified, the vector field $u^a$ should be determined by the intrinsic
geometric properties of the brane.  One can think of several options as to the
choice of this field.  For example, one can identify $u^a$ with a timelike
eigenvector (thereby demanding the existence of such an eigenvector) of a
geometric quantity such as $C^a{}_b$, $R^a{}_b$, $K^a{}_b$, etc.  We will not
attempt to make a specific prescription in this paper. In studying
cosmological scalar perturbations, it will be sufficient to assume that the
unperturbed field $u^a$ is tangent to the world lines of isotropic observers
--- the only distinguished choice --- and that it is perturbed linearly.  In
this case, condition (\ref{bc}) represents a generalization of the approximate
boundary condition derived in \cite{KM,SSH} for the DGP cosmology
\cite{DGP,DDG} (there, the value $\bc = - 1/2$ was obtained).

The specific form (\ref{bc0}) or (\ref{bc}) of our set of boundary conditions,
of course, contains some arbitrariness.  To us it seems to be the simplest one
of the kind of boundary conditions that we propose (specified directly on the
brane); however, we cannot justify its uniqueness at this moment.  Therefore,
it should be regarded as a proposal which demonstrates how the braneworld
theory with our type of boundary conditions can operate in principle, and which
can be tested observationally.  Perhaps, one should keep in mind that the
specific form of boundary conditions in the multi-dimensional space is part of
{\em any\/} braneworld theory with large extra dimension, to be tested against
observations.

After the boundary conditions have been specified, the tensor $C_{ab}$ plays
the role of the stress--energy tensor of an ideal fluid with equation of state
like that of radiation but with a nontrivial dynamics described by
Eq.~(\ref{conserv}). In the literature, this tensor has been called `Weyl
fluid' \cite{W-fluid} and, in the cosmological context, `dark radiation'
\cite{dark-rad}. The stress--energy of this ideal fluid is not conserved due to
the presence of the source term $Q_{ab}$ in (\ref{conserv}). Equation
(\ref{effective}) together with boundary conditions such as those described by
(\ref{weyl})--(\ref{bc}) form a complete set of equations on the brane.

\section{Scalar cosmological perturbations} \label{sec:perturb}

\subsection{Main equations} \label{subsec:main}

The unperturbed metric on the brane is described by the Robertson--Walker line
element and brane expansion is described by \cite{cosmol,DDG,SS02}
\begin{equation} \label{rw}
H^2 + {\kappa \over a^2} = {\rho + \sigma \over 3 m^2} + {2 \over \ell^2}\left[
1 \pm \sqrt{1 + \ell^2 \left({\rho + \sigma \over 3 m^2} - {\Lambda \over 6} -
{C \over a^4} \right)} \right] \, ,
\end{equation}
where $\rho$ is the matter density,
\begin{equation}
\ell = {2 m^2 \over M^3}
\label{ell}
\end{equation}
defines a new fundamental length scale, $\kappa = 0, \pm 1$ describe different
possibilities for the spatial geometry and the term $C/a^4$ ($C$ is a constant)
is the homogeneous contribution from dark radiation. The two signs in
(\ref{rw}) describe two different branches corresponding to the two different
ways in which a brane can be embedded in the Schwarzschild--anti-de~Sitter bulk
\cite{cosmol,DDG}. In \cite{SS02}, we classified models with lower (upper) sign
as Brane\,1 (Brane\,2).  Models with the upper sign can also be called {\em
self-accelerating\/} because they lead to late-time cosmic acceleration even in
the case of zero brane tension and bulk cosmological constant \cite{DDG}.
Throughout this paper, we consider the spatially flat case ($\kappa = 0$) for
simplicity.

On a spatially homogeneous and isotropic brane, the timelike vector field $u^a$
used in the decomposition (\ref{weyl}) coincides with the four-velocity field
of isotropic observers comoving with matter, and the homogeneous dark-radiation
density $\rho_\C$ is related to the constant $C$ by $\rho_\C = - 3 m^2 C /
a^4$. Due to spatial homogeneity and isotropy, we have $\pi_a = 0$ and
$\pi_{ab} = 0$, and equation (\ref{bc}) then implies $\rho_\C = 0 \ \Rightarrow
\ C = 0$ (for $\bc \ne 0$).

Scalar metric perturbations are most conveniently described by the
relativistic potentials $\Phi$ and $\Psi$ in the so-called longitudinal gauge:
\begin{equation}
ds^2 = - ( 1 + 2 \Phi) d t^2 + a^2 ( 1 - 2 \Psi) d \vec{x}^2 \, .
\end{equation}

We denote the components of the linearly perturbed stress--energy tensor of
matter in the coordinate basis as follows:
\begin{equation}
T^\alpha{}_\beta = \left(
\begin{array}{rl}
\displaystyle - (\rho + \delta \rho) \, , & \displaystyle -
\partial_i v
\medskip \\
\displaystyle  {\partial_i v \over a^2} \, , & \displaystyle (p + \delta p)
\delta^i{}_j
\end{array}
\right) \, ,
\end{equation}
where $\delta \rho$, $\delta p$, and $v$ are small quantities.  Similarly, we
introduce the scalar perturbations $\delta \rho_\C$, $v_\C$, and $\delta
\pi_\C$ of the tensor $C_{ab}$ in the coordinate basis:
\begin{equation}
m^2 C^\alpha{}_\beta = \left(
\begin{array}{rl}
\displaystyle {3 m^2 C \over a^4} - \delta \rho_\C
\, , & \displaystyle - \partial_i v_\C \medskip \\
\displaystyle {\partial_i v_\C \over a^2} \, , & \displaystyle \left({\delta
\rho_\C \over 3 } - {m^2 C \over a^4} \right) \delta^i{}_j + {\delta \pi^i{}_j
\over a^2}
\end{array}
\right) \, ,
\end{equation}
where, according to (\ref{decompose}), $\delta \pi_{ij} = \partial_i \partial_j
\delta \pi_\C - \frac13 \delta_{ij} \partial^2 \delta \pi_\C $, and the spatial
indices are raised and lowered with $\delta^i{}_j$.  We call $v$ and $v_\C$ the
momentum potentials for matter and dark radiation, respectively.

Then, equation (\ref{effective}) together with the stress--energy conservation
equation for matter and conservation equation (\ref{conserv}) for dark
radiation result in the following complete system of equations describing the
evolution of scalar perturbations on the brane:
\begin{eqnarray}
\fl \displaystyle \ddot \Psi  + 3(1 + \gamma) H \dot \Psi + H \dot \Phi +
\left[ 2 \dot H + 3 H^2 (1 + \gamma) \right] \Phi - {\gamma \over a^2} \nabla^2
\Psi +
{1 \over 3 a^2} \nabla^2 (\Phi - \Psi) \nonumber \\
= \displaystyle \left[ c_s^2 - \gamma + {2 \over \ell^2 \lambda} \left(c_s^2 -
\frac13 \right) \right] {\delta \rho \over 2 m^2 } + \left( 1 + {2 \over \ell^2
\lambda} \right) {\tau \delta S \over 2 m^2} \, , \label{psi} \\
\fl \delta \dot \rho + 3 H ( \delta \rho + \delta p ) = 3 (\rho + p) \dot \Psi
+ {1 \over a^2} \nabla^2 v \, , \label{rho} \\
\fl \dot v + 3 H v = \delta p + (\rho + p) \Phi \, , \label{v} \\
\fl \delta \dot \rho_\C + 4 H \delta \rho_\C = {1 \over a^2} \nabla^2 v_\C -
{12 m^2 C \over a^4} \dot \Psi \, , \medskip \label{rho-c} \\
\fl \dot v_\C + 3 H v_\C = \frac13 \delta \rho_\C - {4 m^2 C \over a^4} \Phi -
{m^2 \ell^2 \dot \lambda \over 3 H} \left[ {\Delta_\m \over 2 m^2 } - {\nabla^2
\over a^2} \Psi \right] - {m^2 \ell^2 \lambda \over 3 a^2} \nabla^2 (\Phi -
\Psi) \, , \label{v-c}  \\
\fl {1 \over a^2} \nabla^2 \Psi = \left(1 + {2 \over \ell^2 \lambda} \right)
{\Delta_\m \over 2 m^2}  + {\Delta_\C \over m^2 \ell^2 \lambda } \, ,
\label{con-delta} \\
\fl m^2 \ell^2 \lambda \left( \dot \Psi + H \Phi \right) = \left(1 + {\ell^2
\lambda \over 2} \right) v + v_\C \, , \label{con-v} \\
\fl \delta \pi_\C = - {m^2 \ell^2 \lambda \over 4} \left( 1 + 3 \gamma \right)
(\Phi - \Psi) \, . \label{pi}
\end{eqnarray}
Here, we use the following notation: $S$ is the entropy density of the matter
content of the universe, $\tau = \left(
\partial p / \partial S \right)_\rho\,$, $c_s^2 = \left(
\partial p / \partial \rho \right)_S$ is the adiabatic sound velocity, the
time-dependent functions $\lambda$ and $\gamma$ are given by
\begin{equation} \label{lambda}
\lambda \equiv H^2 - {\rho + \sigma \over 3 m^2} - {2 \over \ell^2} = \pm {2
\over \ell^2} \sqrt{1 + \ell^2 \left({\rho + \sigma \over 3 m^2}
- {\Lambda \over 6} - {C \over a^4} \right) } \, ,
\end{equation}
\begin{equation} \label{gamma}
\gamma \equiv \displaystyle \frac13 \left(1 + {\dot \lambda \over H \lambda}
\right) = \frac13 \left[ 1 - {\displaystyle {\rho + p \over m^2} - {4 C \over
a^4} \over \displaystyle 2 \left( {\rho + \sigma \over 3m^2} + {1 \over \ell^2}
- {\Lambda \over 6} - {C \over a^4} \right) } \right] \, ,
\end{equation}
and the perturbations $\Delta_\m$ and $\Delta_\C$ are defined as
\begin{equation}
\Delta_\m = \delta \rho + 3 H v \, , \qquad \Delta_\C = \delta \rho_\C + 3 H
v_\C \, .
\end{equation}
The overdot, as usual, denotes the partial derivative with respect to the time
$t$.

The system of equations (\ref{psi})--(\ref{pi}) generalizes the work of
Deffayet \cite{Deffayet} (for the DGP brane) to the case of a generic
braneworld scenario described by (\ref{action}), which allows non-zero values
for the brane tension and bulk cosmological constant. It describes two
dynamically coupled fluids: matter and dark radiation. It is important to
emphasize that the evolution equations (\ref{rho-c}), (\ref{v-c}) for the
dark-radiation component are {\em not quite\/} the same as those for ordinary
radiation. Of special importance are the last two source terms on the
right-hand side of (\ref{v-c}) which lead to nonconservation of the
dark-radiation density. Thus, the behaviour of this component is rather
nontrivial, as will be demonstrated in the next section.

It is also interesting to note that the perturbations in dark radiation
formally decouple from those in ordinary matter in the important limiting case
$M \to 0$ (equivalently, $\ell \to \infty$), for which the system
(\ref{psi})--(\ref{pi}) reproduces the corresponding equations of general
relativity (after setting $\gamma = c_s^2$).

From equations (\ref{rho})--(\ref{con-v}), one can derive the following useful
system for perturbations in {\em pressureless\/} matter and dark radiation in
the important case $C = 0$\,: {\samepage
\begin{eqnarray}
\ddot \Delta + 2 H \dot \Delta = \left(1 + {6 \gamma \over \ell^2 \lambda}
\right) {\rho \Delta \over 2 m^2} + (1 + 3 \gamma) {\delta\rho_\C \over m^2
\ell^2 \lambda}  \, , \label{one} \\
\dot v_\C + 4 H v_\C = \gamma \Delta_\C + \left(\gamma - \frac13 \right)
\Delta_\m
+ {4 \over 3 (1 + 3 \gamma) a^2 } \nabla^2 \delta \pi_\C \, , \label{two} \\
\delta \dot \rho_\C + 4 H \delta \rho_\C = {1 \over a^2} \nabla^2 v_\C
\label{three} \, ,
\end{eqnarray}
} where
\begin{equation}
\Delta \equiv {\Delta_\m \over \rho}
\end{equation}
is the conventional dimensionless variable describing matter perturbations.

\subsection{Boundary conditions} \label{subsec:bc}

The system of equations (\ref{psi})--(\ref{pi}), or (\ref{one})--(\ref{three}),
describing scalar cosmological perturbations, is not closed on the brane since
the quantity $\delta \pi_\C$ in (\ref{pi}) or (\ref{two}) (hence, the
difference $\Phi - \Psi$) is undetermined and, in principle, can be set
arbitrarily from the brane viewpoint.  This is a particular case of the
nonclosure of the basic equations of the braneworld theory which we noted
earlier.  As discussed in the introduction, our approach to this issue will be
different from that in other papers in that we will not look to the bulk to
specify boundary conditions but, rather, will specify such conditions directly
on the brane.

A general family of boundary conditions on the brane is obtained by relating
the quantities $\pi_\C$ and $\rho_\C$ as discussed in the introduction.
Equation (\ref{bc}) implies $C = 0$ in the case $A \ne 0$ and, in the
linearized form, becomes
\begin{equation} \label{bc-linear}
{1 \over a^2} \nabla^2 \delta \pi_\C  = \bc\, \delta \rho_\C \, .
\end{equation}
In most of this paper, $\bc$ shall be assumed to be a constant. By virtue of
(\ref{pi}), this relates the diffe\-rence $\Phi - \Psi$ between the
gravitational potentials to the perturbation of the dark-radiation density
$\delta \rho_\C$\,:
\begin{equation} \label{phi-psi}
{1 \over a^2} \nabla^2 (\Phi - \Psi) = - {4 \bc \over m^2 \ell^2 \lambda (1 + 3
\gamma)} \delta \rho_\C \, .
\end{equation}

For the boundary condition (\ref{bc-linear}), one can derive a second-order
differential equation for $\delta \rho_\C$ by substituting for $v_\C$ from
(\ref{three}) into (\ref{two}):
\begin{eqnarray} \label{closed}
&{}& \delta \ddot \rho_\C + (10 - \gamma) H \delta \dot \rho_\C + 4 \left[ \dot
H + 3 (2 - \gamma) H^2 \right] \delta \rho_\C  \nonumber \\ &{}& = {1 \over
a^2} \nabla^2 \left[ \gamma \delta \rho_\C + {4 \bc \over 3 (1 + 3 \gamma)}
\delta \rho_\C + \left( \gamma - \frac13 \right) \Delta_\m \right] \, .
\end{eqnarray}
Equations (\ref{one}) and (\ref{closed}) will then form a closed system of two
coupled second-order differential equations for $\Delta$ and $\delta \rho_\C$.
From the form of the right-hand side of (\ref{closed}) one expects this system
to have regions of stability as well as instability. Specifically, a necessary
condition for stability on small spatial scales is that the sign of the
coefficient of $\nabla^2 \delta \rho_\C$ on the right-hand side of
(\ref{closed}) be positive. This leads to the condition\footnote{In the absence
of matter on the brane, $\Delta_\m = 0$, equation (\ref{closed}) becomes a
closed wave-like equation for the scalar mode of gravity, and condition
(\ref{stable}) becomes the boundary of its stability domain.  The existence of
such a scalar gravitational mode is due to the presence of an extra dimension.}
\begin{equation} \label{stable}
\bc \ge - \frac34 \gamma (1 + 3 \gamma) \, .
\end{equation}
From (\ref{gamma}), we find that $\gamma \approx - 1/6$ in a matter-dominated
universe, and condition (\ref{stable}) simplifies to
\begin{equation} \label{stable0}
\bc \ge \frac{1}{16} \, .
\end{equation}

We consider two important subclasses of (\ref{bc-linear}) which we call the
minimal boundary condition and the Koyama--Maartens boundary condition,
respectively:

\subsubsection{Minimal boundary condition.} Our simplest condition
(\ref{bc0}) corresponds to setting $\bc = 0$.  Then, from (\ref{pi}) we obtain
the relation $\Phi = \Psi$, the same as in general relativity.  Under this
condition, equations (\ref{psi})--(\ref{v}) constitute a complete system of
equations for scalar cosmological perturbations on the brane in which initial
conditions for the relativistic potential $\Phi$, $\dot\Phi$ and matter
perturbations $\delta \rho$, $\delta p$, $v$ can be specified quite
independently.  Once a solution of this system is given, one can calculate all
components of dark-radiation perturbations using (\ref{con-delta}) and
(\ref{con-v}). Thus, with this boundary condition, equations
(\ref{rho-c})--(\ref{con-v}) can be regarded as auxiliary and can be used to
felicitate and elucidate the dynamics described by the main system
(\ref{psi})--(\ref{v}). We should stress that only the quantities pertaining to
the induced metric on the brane ($\Phi$, $\Psi$) and those pertaining to matter
($\delta\rho$, $\delta p$, $v$) can be regarded as directly observable, while
those describing dark radiation ($\delta\rho_\C$, $v_\C$) are not directly
observable.

\subsubsection{Koyama--Maartens boundary condition.} In an important paper
\cite{KM}, Koyama and Maartens arrived at condition (\ref{bc-linear}) with
\begin{equation} \label{KM}
\bc = - \frac12 \, .
\end{equation}
This boundary condition was derived in \cite{KM} as an approximate relation in
the DGP model valid only on small (subhorizon) spatial scales under the
assumption of quasi-static behaviour. It was later re-derived in \cite{SSH}
under a similar approximation. However, according to the approach to boundary
conditions taken in the present paper, one can regard (\ref{KM}) as being a
linearized version of the more general relation (\ref{bc}) which is valid on
{\em all\/} spatial scales.  We call it the Koyama--Maartens boundary
condition, although one should be aware of the different status of this
relation in our paper, where it is regarded as an exact additional relation,
and in \cite{KM,SSH}, where it is derived as an approximation to the more
complicated situation.

In discussing the small-scale approximation in quasi-static regime, it was
argued in \cite{KM} that equation (\ref{three}) permits one to neglect the
perturbation $v_\C$ in (\ref{two}), which, together with (\ref{bc-linear}),
will then transform (\ref{one}) into a closed equation for matter
perturbations:\footnote{Equation (\ref{KM-one}) was derived in \cite{KM} only
for the DGP model and for the case $\bc = - 1/2$; however, the argument can be
extended to a general braneworld model and a general value of $\bc$ in
(\ref{bc-linear}).}
\begin{equation} \label{KM-one}
\ddot \Delta + 2 H \dot \Delta = \Theta_{\rm KM}\, {\rho \Delta \over 2 m^2} \,
, \qquad \Theta_{\rm KM} = 1 + { 12 \bc \gamma + 3 \gamma + 1 \over \ell^2
\lambda \left[2 \bc + \frac32 \gamma (1 + 3 \gamma) \right] } \, .
\end{equation}
Some cosmological consequences of this approach are discussed in
\cite{spergel}. As in general relativity, this equation does not contain
spatial derivatives; hence, the evolution of $\Delta$ is independent of the
spatial scale.  We would obtain equation of the type (\ref{KM-one}) for
perturbations had we followed the route of \cite{KM} or \cite{SSH} in finding
approximate solutions of perturbation equations in the bulk and using the
quasi-static approximation (some difficulties with this approach are discussed
in the appendix). However, in the approach adopted in this paper, as will be
shown in the following two sections, numerical integration of the exact
linearized system (\ref{one})--(\ref{three}) does not support approximation
(\ref{KM-one}).  No matter what initial conditions for dark radiation are set
initially, one observes a strong dependence of the evolution of matter
perturbations on the wave number.  In particular, it is incorrect to neglect
the quantity $v_\C$ on small spatial scales, since it is precisely this
quantity which is responsible for the dramatic growth of perturbations both in
$\Delta_\m$ and in $\delta \rho_\C$ on such scales.

From equations (\ref{stable}) and (\ref{stable0}), we can see that the minimal
and Koyama--Maartens boundary conditions generally lead to unstable evolution.
This will be confirmed by numerical simulations in the next section.

\subsubsection{Scale-free boundary conditions.} Evolution of perturbations
in the stability region (\ref{stable}) \& (\ref{stable0}) shows little
dependence on spatial scale. It is interesting that there also exists an
important class of boundary conditions leading to {\em exact\/}
scale-independence. We call these, for simplicity, {\em scale-free boundary
conditions.\/} To remove the dependence on wave number altogether and thereby
obtain a theory in which perturbations in matter qualitatively evolve as in
standard (post-recombination) cosmology, it suffices to set the right-hand side
of (\ref{two}) identically zero:
\begin{equation} \label{bc-new}
{1 \over a^2 } \nabla^2 \delta \pi_\C = {1 + 3 \gamma \over 4} \Bigl[ (1 - 3
\gamma) \Delta_\m - 3 \gamma \Delta_\C \Bigr] \, ,
\end{equation}
which, in view of equation (\ref{con-delta}), can also be expressed in a form
containing only the geometrical quantities $\delta \pi_\C$, $\Delta_\C$, and
$\Psi$.  In this case, the perturbations $\delta v_\C$ and $\delta \rho_\C$ in
dark radiation decay very rapidly, according to equations (\ref{two}) and
(\ref{three}), and (\ref{one}) reduces to the simple equation
\begin{equation} \label{matter}
\ddot \Delta + 2 H \dot \Delta = \Theta\, {\rho \Delta \over 2 m^2} \, , \qquad
\Theta = \left(1 + {6 \gamma \over \ell^2 \lambda} \right) \, ,
\end{equation}
valid on all spatial scales.  Equations (\ref{con-delta}) and (\ref{pi}) then
lead to simple relations between the gravitational potentials $\Phi$ and $\Psi$
and matter perturbations:
\begin{equation} \label{potentials}
{1 \over a^2} \nabla^2 \Phi = \Theta {\Delta_\m \over 2 m^2} \, , \qquad {1
\over a^2} \nabla^2 \Psi = \left(1 + {2 \over \ell^2 \lambda} \right)
{\Delta_\m \over 2 m^2} \, .
\end{equation}
The difference $\Phi - \Psi$ can be conveniently determined from
\begin{equation} \label{difference}
{1 \over a^2} \nabla^2 (\Phi - \Psi) = {3 \gamma - 1 \over m^2 \ell^2 \lambda}
\Delta_\m \, .
\end{equation}

As can easily be seen from (\ref{pi}) or (\ref{difference}), the
general-relativistic relation $\Phi = \Psi$ is not usually valid in braneworld
models. An important exception to this rule is provided by the mimicry models
discussed in section \ref{subsec:scale-free}.

One can propose other conditions of type (\ref{bc-new}) that lead to
scale-independent behaviour. For instance, one can equate to zero the
right-hand side of (\ref{closed}).  But it remains unclear how these may be
generalized to the fully nonlinear case as was done in the case of the boundary
conditions (\ref{bc-linear}) via equation (\ref{bc}). Nevertheless, in view of
the interesting properties of scale-independence and the fact that
perturbations in the stability region (\ref{stable}) \& (\ref{stable0}) behave
in this manner, the consequences of (\ref{matter}) need to be further explored,
and we shall return to this important issue later on in this paper.

Having described the system of linearized equations governing the evolution of
scalar perturbations in pressureless matter and dark radiation, we now proceed
to apply them to two important braneworld models: the popular DGP model
\cite{DGP,DDG} and the `mimicry' model suggested in \cite{SSV}. It should be
noted that these two models are complementary in the sense that the mimicry
model arises for {\em large values\/} of the bulk cosmological constant
$\Lambda$ and brane tension $\sigma$, whereas the DGP cosmology corresponds to
the opposite situation $\Lambda = 0$ and $\sigma = 0$.

\section{Scalar perturbations in the DGP model} \label{sec:DGP}

Amongst alternatives to LCDM, the Dvali--Gabadadze--Porrati (DGP) model
\cite{DGP} stands out because of its stark simplicity. Like the cosmological
constant which features in LCDM, the DGP model too has an extra parameter $\ell
= {2 m^2 / M^3}$, the length scale beyond which gravity effectively becomes
five-dimensional. However, unlike the cosmological constant whose value must be
extremely small in order to satisfy observations, the value $\ell \sim
cH_0^{-1}$, required to explain cosmic acceleration, can be obtained by a
`reasonable' value of the five-dimensional Planck mass $M \sim 10$~MeV. As
pointed out earlier, DGP cosmology belongs to the class of induced gravity
models which we examine and is obtained from (\ref{action}) after setting to
zero the brane tension and the cosmological constant in the bulk (i.e., $\sigma
= 0$ and $\Lambda=0$). Under the additional assumption of spatial flatness
($\kappa = 0$) and $C = 0$, the modified Friedmann equation (\ref{rw}) becomes
\cite{DDG}
\begin{equation} \label{friedman-DGP}
H^2 - \frac{2H}{\ell} = \frac{\rho}{3m^2}  \, .
\end{equation}
In a spatially flat universe, given the current value of the matter density and
Hubble constant, $\ell$ ceases to be a free parameter and becomes related to
the matter density by the following relation
\begin{equation}
\Omega_\ell \equiv {1 \over \ell^2 H_0^2} = \left (\frac{1-\Omega_\m}{2}\right
)^2 \, ,
\end{equation}
which may be contrasted with $\Omega_\Lambda = 1 - \Omega_\m$ in LCDM.

\begin{figure}[tbh!]
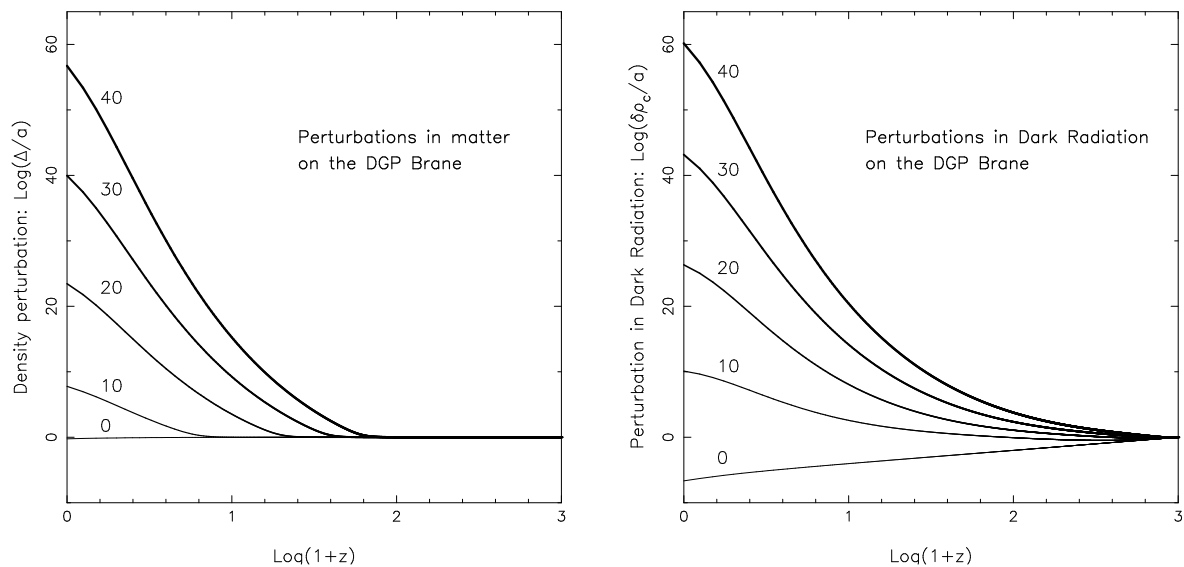

\centerline{ \psfig{figure=dgp_KM1-y4=1.ps,width=0.47\textwidth,angle=0} \ \ \
\ \ \psfig{figure=dgp_KM-DR-y4=1.ps,width=0.47\textwidth,angle=0} }
\caption{\small {\bf The DGP brane with the Koyama--Maartens boundary condition
$\bc = - 1/2$.} Growth of perturbations in matter (left) and dark radiation
(right) on the DGP brane are shown for different values of the comoving wave
number $k/a_0 H_0$ (indicated by numbers above the corresponding curves). The
current matter density is chosen to be $\Omega_\m = 0.22$, and the initial
value of $v_\C$ is set to zero. (Our results remain qualitatively the same for
other values of the density parameter.) Note the dramatic $k$-dependence in the
growth of perturbations both in matter and in dark radiation. For comparison,
$\Delta / a = 1$ at these redshifts in the standard CDM model with $\Omega_\m =
1$. \label{fig:dgp}}
\end{figure}

Linear perturbation equations for this model were discussed in
\cite{LSS,KM,Mukohyama,Deffayet}. An approximate boundary condition for scalar
perturbations was obtained by Koyama and Maartens \cite{KM} on subhorizon
scales, and it is described by equations (\ref{bc-linear}), (\ref{KM}).  For
convenience, we present system (\ref{one})--(\ref{three}) for this
case\footnote{To facilitate comparison, we relate the notation of
Ref.~\cite{KM} with that of our paper. Our length scale $\ell$ is related with
the length scale $r_c$ of \cite{KM} by $\ell = 2 r_c$.  Our quantity $\delta
\rho$ is the same as in \cite{KM}, and our quantities $v$, $\delta \rho_\C$,
$v_\C$, and $\delta \pi_\C$ in the DGP model are related to the similar
quantities $\delta q$, $\delta \rho_E$, $\delta q_E$, and $\delta \pi_E$ of
\cite{KM} by the equalities $v = - a \delta q$, $\delta \rho_\C = - \delta
\rho_E$, $v_\C = a \delta q_E$, and $\delta \pi_\C = - a^2 \delta \pi_E$.}:
\begin{eqnarray}
\ddot \Delta + 2 H \dot \Delta = \left(1 + {6 \gamma \over \ell^2 \lambda}
\right) {\rho \Delta \over 2 m^2} + (1 + 3 \gamma) {\delta\rho_\C \over m^2
\ell^2 \lambda}  \, , \label{DGP-one} \\
\dot v_\C + 4 H v_\C = \gamma \Delta_\C + \left(\gamma - \frac13 \right)
\Delta_\m
- {2 \over 3 (1 + 3 \gamma) } \delta \rho_\C \, , \label{DGP-two} \\
\delta \dot \rho_\C + 4 H \delta \rho_\C = {1 \over a^2} \nabla^2 v_\C
\label{DGP-three} \, .
\end{eqnarray}
In the DGP model, the general expressions (\ref{lambda}) and (\ref{gamma}) for
$\lambda$ and $\gamma$ in the case of pressureless matter reduce to
\begin{equation} \label{DGP-lagam}
\lambda = {2 \over \ell} \left(H - \frac1\ell \right)\, , \qquad \gamma = {1
\over 2 (\ell H - 1)^2} - \frac16 \, .
\end{equation}

\begin{figure}[tbh!]
\centerline{\psfig{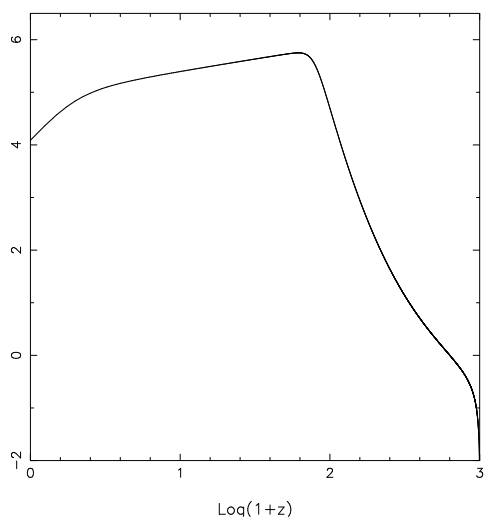}
} \caption{\small Shown is the logarithm of the ratio (\ref{ratio}) for scalar
perturbations in the DGP model described by the system of equations
(\ref{DGP-one})--(\ref{DGP-three}) with comoving wave number $k/a_0 H_0 = 50$.
Initial conditions (at $z = 10^3$) are chosen such that $v_\C = 0$ and $\dot
v_\C = 0$, so that the initial value of the ratio (\ref{ratio}) is equal to
zero. Observe that this ratio grows quickly and exceeds unity by several orders
of magnitude at late times (low redshifts) demonstrating the breakdown of the
quasi-static approximation.
\label{fig:ratio}}
\end{figure}

\begin{figure}[tbh!]
\centerline{\psfig{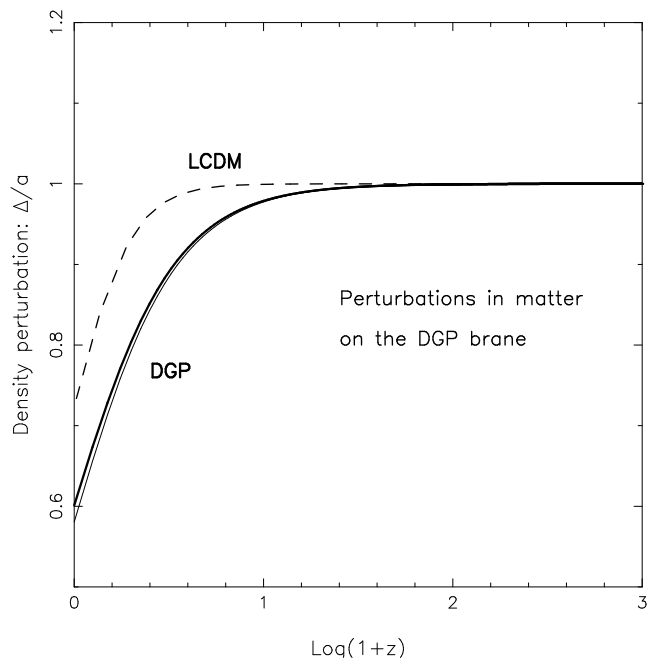} }
\caption{\small {\bf The DGP brane with the boundary condition $\bc = 1/2$.}
The parameters of the model are the same as in figure~\ref{fig:dgp} but the
y-axis is no longer plotted in logarithmic units. The two solid curves show the
evolution of scalar perturbations corresponding to the comoving wave numbers $k
/ a_0 H_0 = 0$ (thin curve) and $40$ (thick curve). Note that these two curves
are almost indistinguishable which illustrates that the growth of perturbations
is virtually scale-independent in this case. The dashed line shows the
behaviour of scalar perturbations in the LCDM model. (In all cases $\Omega_\m =
0.22$ is assumed.) \label{fig:dgp-stable}}
\end{figure}

The results of a typical integration of the exact system of equations
(\ref{DGP-one})--(\ref{DGP-three}) for different values of the wave number $k$
are shown in figure~\ref{fig:dgp}. We observe a dramatic escalation in the
growth of perturbations at moderate redshifts and a strong $k$-dependence for
perturbations in matter as well as in dark-radiation (the y-axis is plotted in
logarithmic units). These results do not support the approximation made in
\cite{KM}, which assumes the left-hand side of equation (\ref{DGP-two}) to be
much smaller than individual terms on its right-hand side for sufficiently
large values of $k$, and which leads, subsequently, to the scale-independent
equation (\ref{KM-one}).  To show explicitly that this `quasi-static' regime is
rather unlikely, we have integrated the system
(\ref{DGP-one})--(\ref{DGP-three}) for a sufficiently high comoving value
$k/a_0 H_0 = 50$ and with the initial condition $v_\C = 0$ and $\dot v_\C = 0$,
which sets both sides of equation (\ref{DGP-two}) initially to zero.  In
figure~\ref{fig:ratio}, we plot the logarithm of the ratio
\begin{equation} \label{ratio}
\left|  { \dot v_\C + 4 H v_\C \over \left( \gamma - \frac13 \right) \Delta_\m
} \right| \, ,
\end{equation}
which is assumed to be $\ll 1$ in the quasi-static approximation of \cite{KM}.
From figure~\ref{fig:ratio} we see that this ratio soon becomes much larger
than unity. This demonstrates that, even if $v_\C$ and $\dot v_\C$ were small
initially, during the course of expansion both quantities grow to large values
casting doubt on the validity of the quasi-static approximation for
small-wavelength modes.

We would like to stress that our conclusions themselves are not based on the
small-scale or quasi-static approximation.  Indeed, we integrate the {\em
exact\/} system of equations (\ref{one})--(\ref{three}) on the brane, and the
only ansatz that we set in this system is the boundary condition
(\ref{bc-linear}), (\ref{KM}).  Our failure to find the regime in which the
ratio (\ref{ratio}) remains small indicates that such regime is incompatible
with the boundary condition (\ref{bc-linear}), (\ref{KM}).

\begin{figure}[tbh!]
\centerline{\psfig{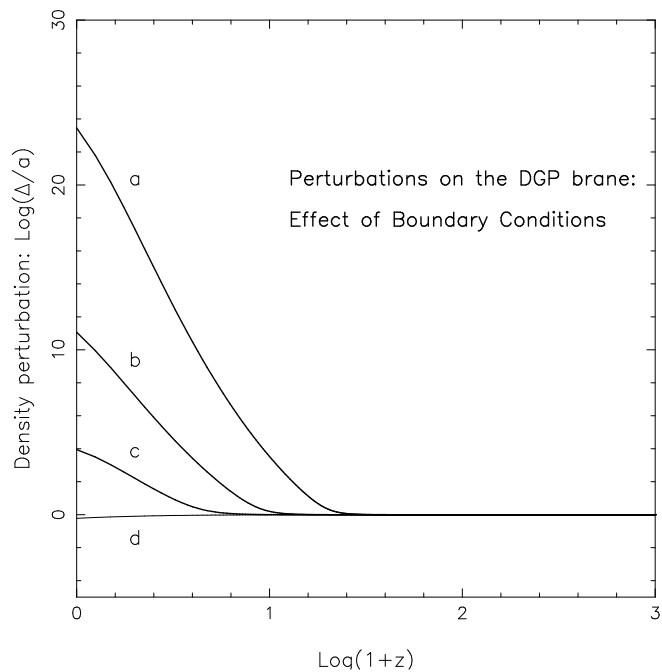} }
\caption{\small Growth of scalar perturbations in matter on the DGP brane is
shown for different boundary conditions in the brane--bulk system. (The
comoving wave number $k/a_0 H_0 = 20$ in all cases.) Boundary conditions are
specified by (\ref{bc-linear}) and differ in the expression for $\bc$; namely:
(a)~the Koyama--Maartens condition $\bc = - 1/2$, (b)~$\bc = - (1 + 3 \gamma) /
4$, (c)~the minimal condition $\bc = 0$, and (d)~$\bc = 1/2$. Condition (b)
with time-dependent value of $\bc$ was chosen because it simplifies equation
(\ref{v-c}), and condition (d) because it lies well inside the stability domain
(\ref{stable}). For comparison,  note that $\Delta / a = 1$ for all values of
$k$ in the standard CDM model with $\Omega_\m = 1$. \label{fig:bc}}
\end{figure}

The strong $k$-dependence of the evolution of perturbations can be explained by
the presence of the term $\nabla^2 v_\C$ on the right-hand side of
(\ref{DGP-three}), which leads to the generation of large perturbations of dark
radiation $\delta \rho_\C$.  The quantity $v_\C$ is being generated by the
right-hand side of equation (\ref{DGP-two}).  The instability in the growth of
perturbations for the Koyama--Maartens boundary condition is in agreement with
the fact that the value of $\bc = - 1/2$ lies well beyond the stability domain
(\ref{stable0}).

As demonstrated earlier, depending upon the value of $\bc$, perturbations on
the brane can be either unstable or quasi-stable. By unstable is meant
$\Delta/a \gg 1$ while quasi-stability implies $\Delta/a \sim O(1)$. The
quasi-stable region (\ref{stable}) is illustrated in
figure~\ref{fig:dgp-stable}, in which we show the results of a numerical
integration of equations (\ref{one})--(\ref{three}) for $\bc = 1/2$.  It is
instructive to compare this figure with the left panel of figure~\ref{fig:dgp}.
One clearly sees the much weaker growth of perturbations as well as their
scale-independence in this case. We therefore conclude that boundary conditions
can strongly influence the evolution of perturbations on the brane. Our results
are summarized in figure~\ref{fig:bc}, which shows the evolution $\Delta/a$
obtained by integrating the system (\ref{one})--(\ref{three}) for different
boundary conditions. (Results for the wave number $k/a_0 H_0 = 20$ are shown.)
We see that the growth of perturbations becomes weaker as the value of $\bc$
approaches the stability domain (\ref{stable}), and quasi-stability is observed
for $\bc = 1/2$.

The behaviour of scalar perturbations on the DGP brane in the case of
scale-free boundary conditions (\ref{bc-new}) is very similar to that shown in
figure~\ref{fig:dgp-stable}. As in \cite{LSS,KM}, we also find that
perturbation growth on the DGP brane is slower than that in LCDM.

\section{Scalar perturbations in mimicry models} \label{sec:mimicry}

\subsection{Cosmic mimicry on the brane}

In our paper \cite{SSV}, we described a braneworld model in which cosmological
evolution proceeds similarly to that of the Friedmannian cosmology but with
different values of the effective matter parameter $\Omega_{\rm m}$, or,
equivalently, with different values of the effective gravitational constant
$G_{\rm N}$, at different cosmological epochs.

Specifically, for a spatially flat universe with zero background dark radiation
($C = 0$), the cosmological equation (\ref{rw}) can be written as follows:
\begin{equation} \label{mimic}
H^2 = {\Lambda \over 6} + \left[ \sqrt{ {\rho - \rho_0 \over 3 m^2} + \left(
\sqrt{H_0^2 - {\Lambda \over 6}} \mp \frac1\ell \right)^2 } \pm \frac1\ell
\right]^2 \, ,
\end{equation}
where $\rho_0$ and $H_0$ are the energy density of matter and Hubble parameter,
respectively, at the present moment of time.  Proceeding from (\ref{rw}) to
this equation, we traded the brane tension $\sigma$ for the new parameters
$H_0$ and $\rho_0$.

In the effective mimicry scenario \cite{SSV}, the parameters $H_0^2 - \Lambda /
6$ and $1 / \ell^2$ are assumed to be of the same order, and much larger than
the present matter-density term $\rho_0 / 3 m^2$. The mimicry model has two
regimes: one in the deep past (high matter density) and another in the more
recent past (lower matter density).  In the deep past, we have
\begin{equation}
{\rho - \rho_0 \over 3 m^2} \gg \left(\sqrt{H_0^2 - {\Lambda \over 6}} \mp {1
\over \ell} \right)^2 \, ,
\end{equation}
and the universe expands in a Friedmannian way
\begin{equation} \label{past}
H^2 \approx {\rho \over 3 m^2} \, .
\end{equation}

In the recent past, we have
\begin{equation}
{\rho - \rho_0 \over 3 m^2} \ll \left(\sqrt{H_0^2 - {\Lambda \over 6}} \mp {1
\over \ell} \right)^2 \, ,
\end{equation}
and the expansion law is approximated by
\begin{equation} \label{future}
H^2 \approx H_0^2 + {\alpha \over \alpha \mp 1} {\rho - \rho_0 \over 3 m^2} \,
,
\end{equation}
where
\begin{equation}
\alpha = \ell \sqrt{H_0^2 - {\Lambda \over 6}}
\end{equation}
is the parameter introduced in \cite{SSV}.  In the case of the cosmological
branch with upper sign, it is assumed that the coefficient $\alpha / (\alpha -
1)$ in (\ref{future}) and in similar expressions is always positive, i.e.,
$\alpha$ is assumed to be greater than unity in this case.

One can interpret the result (\ref{future}) either as a renormalization of the
effective gravitational constant relative to its value in the deep past or as a
renormalization of the effective density parameter:
\begin{equation} \label{lcdm}
H^2 \approx H_0^2 + {\rho^{\rm LCDM} - \rho^{\rm LCDM}_0 \over 3 m^2} \, ,
\qquad \rho^{\rm LCDM} = {\alpha \over \alpha \mp 1} \rho \, .
\end{equation}

In a universe in which matter dominates in the energy density $\rho$,
introducing the cosmological parameters
\begin{equation}
\Omega_\m = {\rho_0 \over 3 m^2 H_0^2} \, ,  \quad \Omega_\ell = {1 \over
\ell^2 H_0^2} \, , \quad \Omega_\Lambda = - {\Lambda \over 6 H_0^2} \, ,
\end{equation}
one can express the Hubble parameter as a function of redshift for the case
under consideration:
\begin{eqnarray}
{H^2(z) \over H_0^2} &=& \Omega_{\rm m} (1\!+\!z)^3 + 1 -  \Omega_{\rm m} +
2 \Omega_\ell \mp 2\sqrt{\Omega_\ell}\, \sqrt{1+\Omega_\Lambda}\nonumber \\
&\pm& 2 \sqrt{\Omega_\ell}\, \sqrt{\Omega_{\rm m} (1\!+\!z)^3 - \Omega_{\rm m}
+ \left( \sqrt{1+\Omega_\Lambda} \mp \sqrt{\Omega_\ell} \right)^2} \, .
\label{hubble1}
\end{eqnarray}

For large values of the extra-dimensional parameters $\Omega_\Lambda$ and
$\Omega_\ell\,$, for which the domain of redshifts $z$ exists such that
\begin{equation}
\Omega_{\rm m} (1\!+\!z)^3 \ll \left( \sqrt{1+\Omega_\Lambda} \mp
\sqrt{\Omega_\ell} \right)^2 \, , \label{eq:mimic0}
\end{equation}
expression (\ref{hubble1}) in this domain reduces to
\begin{equation}
{H^2(z) \over H_0^2} \approx \Omega^{\rm LCDM}_{\rm m} (1\!+\!z)^3 + 1 -
\Omega^{\rm LCDM}_{\rm m}\, , \label{eq:lcdm}
\end{equation}
where
\begin{equation} \label{omega-lcdm}
\Omega^{\rm LCDM}_{\rm m} = {\alpha \over \alpha \mp 1}\, \Omega_{\rm m}  \, ,
\quad \alpha = {\sqrt{1 + \Omega_\Lambda} \over \sqrt{\Omega_\ell}} \, .
\end{equation}
Equation (\ref{eq:lcdm}) is equivalent to (\ref{future}).

\begin{figure}[tbh!]
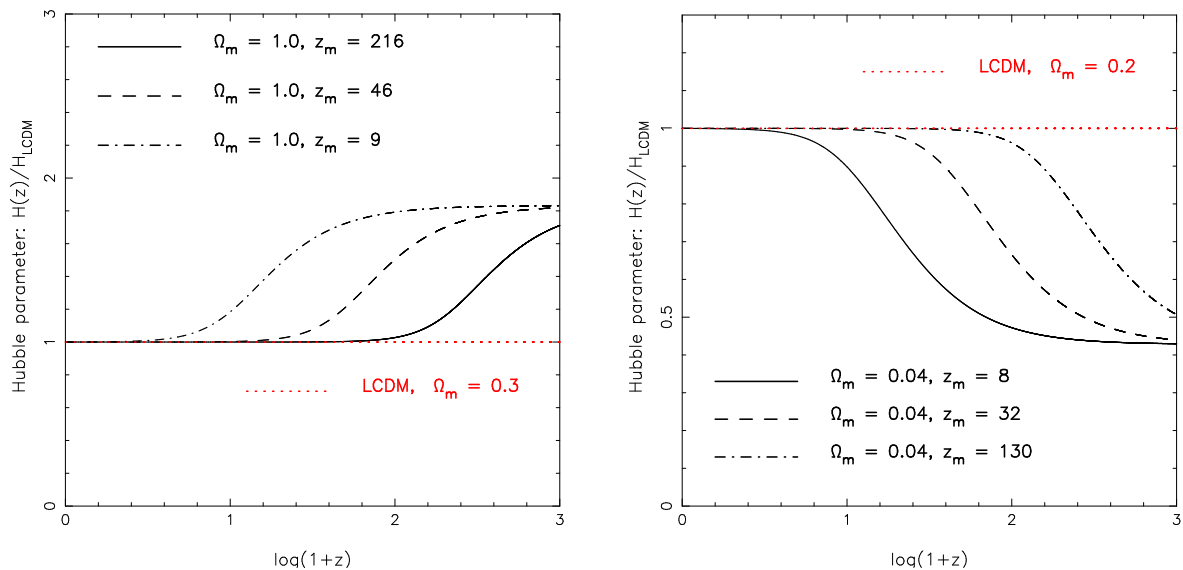

\centerline{\psfig{figure=b1.ps,width=0.47\textwidth,angle=-90} \ \ \ \ \
\psfig{figure=b2.ps,width=0.47\textwidth,angle=-90} } \caption{\small This
figure illustrates that $H_{\rm brane} \simeq H_{\rm LCDM}$ for low redshifts,
whereas at higher redshifts $H_{\rm brane} > H_{\rm LCDM}$ (for Mimicry\,1,
shown in the left panel) or $H_{\rm brane} < H_{\rm LCDM}$ (for Mimicry\,2,
shown in the right panel). This behaviour arises when $\Omega_\m >
\Omega_\m^{\rm LCDM}$ (Mimicry\,1) or $\Omega_\m < \Omega_\m^{\rm LCDM}$
(Mimicry\,2). See \cite{SSV} for details of mimicry models and for some
cosmological consequences of mimicry.\label{fig:mimicry}}
\end{figure}

Remarkably, the behaviour of the Hubble parameter on the brane practically {\em
coincides\/} with that in LCDM at low redshifts $z \ll z_{\m}$, where
\begin{equation} \label{eq:mimic1}
z_{\m} = \frac{ \left(\sqrt{1 + \Omega_\Lambda} \mp \sqrt{\Omega_\ell}
\right)^{2/3}} {\Omega_{\rm m}^{1/3}} - 1 \, .
\end{equation}
This property was called {\em `cosmic mimicry'\/} in \cite{SSV} and $z_\m$ the
{\em mimicry redshift\/}.  An important aspect of the mimicry model,
illustrated by (\ref{eq:lcdm}) and (\ref{omega-lcdm}), is that the matter
density entering (\ref{eq:lcdm}) is an {\em effective\/} quantity. A
consequence of this is the fact that the two density parameters $\Omega_{\rm
m}$ and $\Omega^{\rm LCDM}_{\rm m}$ need not be equal and, based solely on
observations of the coordinate distance, a low/high density braneworld could
easily masquerade as LCDM with a moderate value $\Omega^{\rm LCDM}_{\rm m}
\simeq 0.2 - 0.3$. This situation is illustrated in figure~\ref{fig:mimicry}.

We remind the reader that the two signs in (\ref{mimic}) or (\ref{hubble1})
correspond to two complementary possibilities for embedding the brane in the
higher-dimensional (Schwarzschild-AdS) bulk. In our ensuing discussion, we
shall refer to the model with the lower (upper) sign in (\ref{hubble1}) as
Mimicry\,1 (Mimicry\,2), and we consider these two models separately.

\subsection{Cosmological perturbations in the mimicry model: \\
I.~Scale-dependent boundary conditions}

As expected, perturbations in the mimicry model crucially depend upon the type
of boundary condition which has been imposed. Generally speaking, brane
perturbations grow moderately for BC's which lie in the stability domain
(\ref{stable}) or (\ref{stable0}) and more rapidly in the instability region.
This remains true for mimicry models. In this section, we explore the behaviour
of perturbations in this model for the boundary condition $\bc = 0$, which
belongs to the instability class. In the next section, we shall explore BC's
which give rise to more moderate and scale-independent behavior.

\begin{figure}[tbh!]
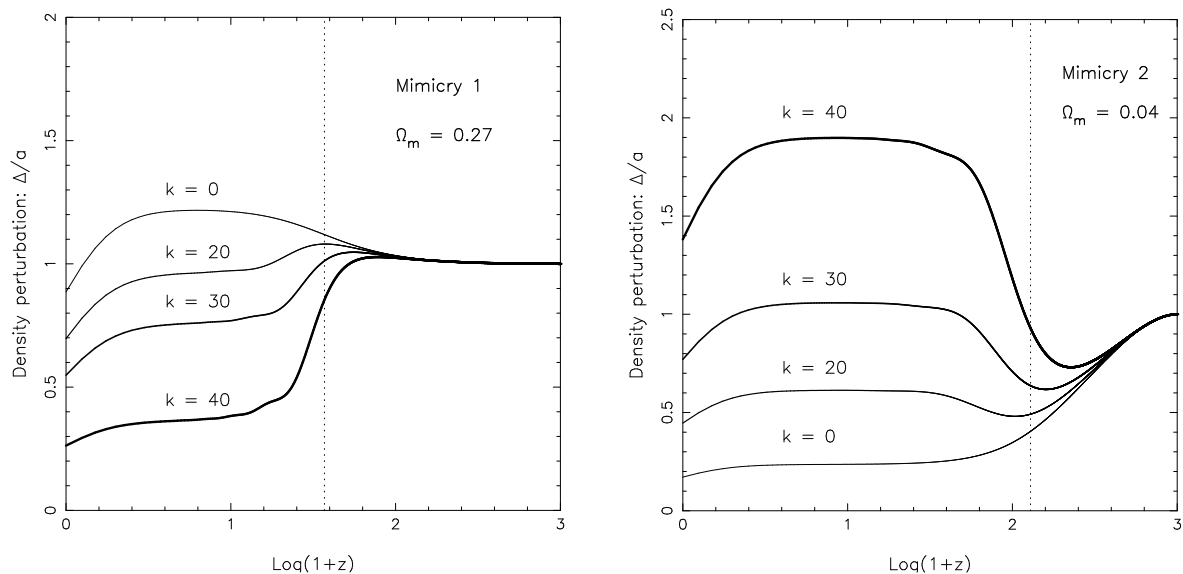

\centerline{ \psfig{figure=M1-S=0.ps,width=0.47\textwidth,angle=0} \ \ \ \ \
\psfig{figure=M2-S=0.ps,width=0.47\textwidth,angle=0} } \caption{\small Growth
of matter perturbations in the Mimicry\,1 and Mimicry\,2 models are shown for
different values of the comoving wave number $k/a_0 H_0$ (indicated by numbers
above the corresponding curves) and for the minimal boundary condition $\bc =
0$ in (\ref{bc-linear}). Both models have the same effective parameter
$\Omega^{\rm LCDM}_\m = 0.22$, hence, quite different matter content, indicated
by the parameter $\Omega_\m$. The value of $v_\C$ is set to zero initially. The
position of the mimicry redshift $z_\m$ is indicated by the vertical dotted
lines. \label{fig:mimicry12}}
\end{figure}

The growth of perturbations if $\bc = 0$ is substituted in (\ref{bc-linear}) is
illustrated in figure~\ref{fig:mimicry12}. The $k$-dependence, clearly seen in
this figure, can be understood by inspecting the system of equations
(\ref{one})--(\ref{three}). Even if we start with zero initial conditions for
the dark-radiation components $\delta \rho_\C$ and $v_\C$, the nontrivial
right-hand side of Eq.~(\ref{two}) leads to the generation of $v_\C$; then, via
the $k$-dependent right-hand side of (\ref{three}), the density $\delta
\rho_\C$ is generated, which later influences the growth of perturbations of
matter in (\ref{one}).  The instability in the growth of perturbations is
explained by the fact that the value of $\bc = 0$ lies outside the stability
domain (\ref{stable}) or (\ref{stable0}).  However, the growth of perturbations
is not as dramatic in this case as in the DGP model with the Koyama--Maartens
BC's, mainly because the value $\bc = 0$ lies much closer to the boundary
(\ref{stable0}) than the Koyama--Maartens value $\bc = - 1/2$.

Qualitatively, the evolution of matter perturbations in mimicry models can be
understood as follows: during the early stages of matter-domination the last
term on the right-hand side of equation (\ref{one}) is not very important,
which transforms (\ref{one}) into a closed equation for the matter
perturbation. Indeed, in the pre-mimicry regime, for $z \gg z_\m\,$, we have
$\left| \ell^2 \lambda \right| \gg 1$ for the quantity in the denominator of
the last term on the right-hand side of (\ref{one}), which makes this term
relatively small for moderate values of $\delta \rho_\C$. Thus, perturbations
in matter evolve according to (\ref{matter}) on all spatial scales, for
redshifts greater than the mimicry redshift $z_\m$. For $z \leq z_\m$, the
quantity $\left| \ell^2 \lambda \right|$ is of order unity. By this time, the
perturbations $\delta \rho_\C$ have grown large, and their amplitude strongly
depends on the wave number. Through the last term in equation (\ref{one}), they
begin to influence the growth of matter perturbations for $z \sim z_\m$,
resulting in the $k$-dependent growth of the latter. The reason for the
opposite $k$-dependence of matter perturbations in Mimicry\,1 and Mimicry\,2
shown in figure~\ref{fig:mimicry12} is connected with the difference in the
sign of $\lambda$ --- defined in (\ref{gamma}) --- for the two models.  Thus,
the last term in (\ref{one}) comes with opposite signs in Mimicry\,1 and
Mimicry\,2, and therefore works in opposite directions in these two models.

Well inside the mimicry regime, for $z \ll z_\m$, we have $\gamma \approx 1/3$,
so that the second term on the right-hand side of (\ref{two}) can be ignored if
matter perturbations are not too large. Then equations (\ref{two}),
(\ref{three}), and (\ref{bc-linear}) lead to a closed system of equations for
the evolution of dark-radiation perturbations. Substituting $\gamma = 1/3$ into
this system, we obtain:
\begin{equation} \label{vc}
v_\C = a^{- 7/2} \xi \, , \qquad \delta \rho_\C = {3 \over (1 + 2 \bc) a^3}
{\partial \over
\partial t} \left(a^3 v_\C \right) \, ,
\end{equation}
where we assumed $\bc \not \approx - 1/2$ to be constant. The function $\xi$
obeys an oscillator-type equation
\begin{equation} \label{xi}
\ddot \xi - \left( \frac12 \dot H + \frac14 H^2 + {1 + 2 \bc \over 3 a^2}
\nabla^2 \right) \xi = 0 \, .
\end{equation}
This means that both $\delta\rho_\C$ and $v_\C$ rapidly decay during the
mimicry regime (oscillating approximately in opposite phase) and the last term
on the right-hand side of (\ref{one}) again becomes unimportant.  In
particular, this will describe the behaviour of the mimicry model with the
minimal boundary condition $\bc = 0$.  The transient oscillatory character of
$\delta \rho_\C$ induces transient oscillations with small amplitude in
$\Delta$ through the last term in (\ref{three}). These small oscillations can
be noticed in figure~\ref{fig:mimicry12} for $\log\, (1 + z) \gsim 1$,
particularly for values $40$ and $30$ of the comoving wave number.\footnote{For
the Koyama--Maartens boundary condition $\bc = - 1/2$, the approximation
described above is not valid during the mimicry stage.  Instead, during
mimicry, the value of $v_\C$ decays without oscillating approximately as $v_\C
\propto 1/a^3$, as can be seen from equations (\ref{two}), (\ref{bc-linear}),
and the value of $\delta \rho_\C$ also decays, which follows from
(\ref{three}).}

Two important features of mimicry models deserve to be highlighted:
\begin{enumerate}

\item As demonstrated in figure~\ref{fig:mimicry12}, there is a strong
suppression of long-wavelength modes in Mimicry\,2.

\item From this figure, we also find that the growth of short-wavelength modes
in Mimicry\,2 can be substantial, even in a low-density universe.

\end{enumerate}

Both properties could lead to interesting cosmological consequences. For
instance, the relative suppression of low-$k$ modes may lead to a corresponding
suppression of low-multipole fluctuations in the cosmic microwave background,
while the increased amplitude of high-$k$ modes could lead to an earlier epoch
of structure formation. (Since the mimicry models behave as LCDM at low
redshifts, they satisfy the supernova constraints quite well.) A detailed
investigation of both effects, however, requires that we know the form of the
transfer function of fluctuations in matter (and dark radiation) at the end of
the radiative epoch. Such an investigation lies outside the scope of the
present paper, but we may return to it elsewhere.\footnote{For simplicity, the
amplitudes of all $k$-modes were assumed to be equal at high redshifts in
figures \ref{fig:dgp}, \ref{fig:bc} and \ref{fig:mimicry12}. A more realistic
portrayal of $\Delta(k)$ should take into consideration the initial spectrum
and the properties of the transfer function for matter and dark radiation, and
we shall return to this in a future work.}

For the minimal boundary condition ($\bc = 0$), assumed in this section,
equations (\ref{con-delta}) and (\ref{phi-psi}) imply $\Phi = \Psi$ and
\begin{equation}
{1 \over a^2} \nabla^2 \Phi = \left(1 + {2 \over \ell^2 \lambda} \right)
{\Delta_\m \over 2 m^2}  + {\Delta_\C \over m^2 \ell^2 \lambda } \, ,
\end{equation}
which is a generalization of the Poisson equation for the mimicry brane.

Mimicry models with the Koyama--Maartens boundary condition exhibit much
stronger instability in the growth of $\delta \rho_\C$ for high values of $k$,
enhancing the growth of matter perturbations (not shown).  This can be
explained by the fact that $\bc = 0$ is much closer to the boundary of the
stability domain (\ref{stable0}) than the Koyama--Maartens value $\bc = - 1/2$.
For the latter, the density perturbation $\Delta$ in the Mimicry\,1 model grows
to be large and negative, while the perturbation $\delta \rho_\C$ becomes large
and positive; for instance, both $\Delta$ and the dimensionless quantity
$\delta \rho_\C / m^2 H^2$ grow by a factor of $10^{11}$ for $k / a_0 H_0 =
40$. This provides another example of the very strong dependence of
perturbation evolution on boundary conditions. 

In our calculations, we have not found any significant dependence of the
eventual growth of perturbations on initial conditions for dark radiation
specified in a reasonable range (at $z = 10^3$).

\subsection{Cosmological perturbations in the Mimicry model: \\
II.~Scale-free boundary conditions} \label{subsec:scale-free}

As mentioned earlier, BC's lying in the stability region (\ref{stable}) lead to
an almost scale-free growth of density perturbations. A similar result is
obtained if we assume the scale-free boundary condition (\ref{bc-new}) of
section \ref{subsec:bc}. In this case, the momentum potential $v_\C$ decays as
$v_\C \propto a^{-4}$, and its spatial gradients in (\ref{three}) can therefore
be neglected. The same is true, of course, if one considers super-horizon modes
with $k \ll aH$. In both cases, we have approximately $\delta\rho_\C \propto
1/a^4$, suggesting that the dynamical role of perturbations in dark radiation
is unimportant. This results in a radical simplification: as in the DGP model,
for BC's lying in the stability region, the growth of perturbations in matter
can be effectively described by a single second-order differential equation
(\ref{matter}), namely,
\begin{equation} \label{matter1}
\ddot \Delta + 2 H \dot \Delta = \Theta\, {\rho \Delta \over 2 m^2} \, , \qquad
\Theta = \left(1 + {6 \gamma \over \ell^2 \lambda} \right) \, ,
\end{equation}
where $\lambda$ and $\gamma$ are defined in (\ref{lambda}) and (\ref{gamma}),
respectively. We shall call $\Theta(z)$ in (\ref{matter1}) the {\em `gravity
term'\/} since it incorporates the effects of modified gravity on the growth of
perturbations. The value of this term on the brane can depart from the
canonical $\Theta = 1$ in general relativity.

\begin{figure}[tbh!]
\begin{center}
\centerline{\psfig{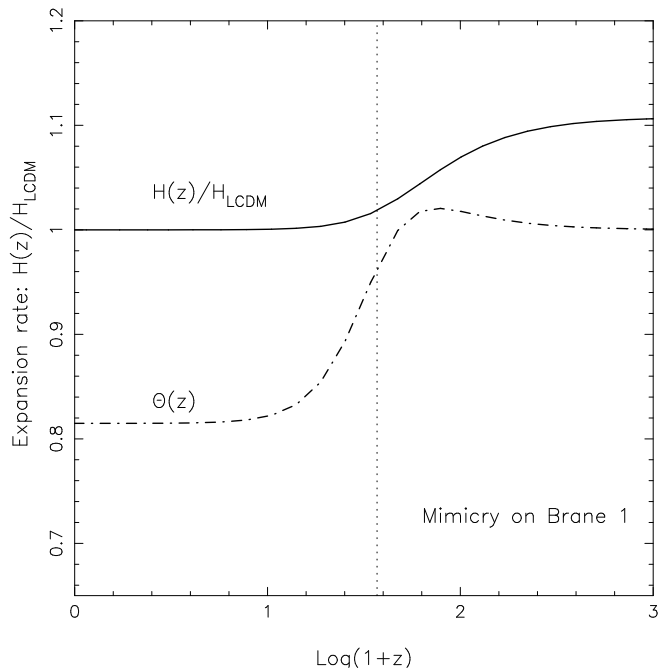} }
\end{center}
\caption{\small The Hubble parameter in the braneworld `Mimicry~1' is shown
relative to LCDM (solid). The LCDM model has $\Omega^{\rm LCDM}_{\rm m} = 0.22$
while $\Omega_{\rm m} = 0.27$ on the brane. Also shown is the `gravity term'
$\Theta(z)$ defined in (\ref{matter}) whose value diminishes from unity at high
redshifts to the asymptotic form (\ref{theta1}) at low redshifts. The dotted
vertical line shows the mimicry redshift $z_m \approx 37$.} \label{fig:theta}
\end{figure}

Figure~\ref{fig:theta} shows the behaviour of $\Theta(z)$ for a typical mimicry
model. At redshifts significantly larger than the mimicry redshift, $z \gg
z_\m$, we have $\Theta(z) \simeq 1$, whereas at low redshifts, $z \ll z_\m$,
the value of $\Theta(z)$ changes to
\begin{equation} \label{theta1}
\Theta(z) \simeq \frac{\Omega^{\rm LCDM}_\m}{\Omega_\m} = \frac{\rho^{\rm
LCDM}}{{\rho}} \quad \mbox{for} \quad z \ll z_\m \, ,
\end{equation}
where $\rho^{\rm LCDM}$ is defined in (\ref{lcdm}).  The solid line in the same
figure shows the ratio of the Hubble parameter on the brane to that in LCDM.
The consequences of this behaviour for the growth equation (\ref{matter}) are
very interesting. Substituting (\ref{theta1}) into (\ref{matter}) and noting
that $H(z) \simeq H^{\rm LCDM}$ during mimicry, we recover the standard
equation describing perturbation growth in the LCDM model
\begin{equation}
{\ddot \Delta} + 2 H^{\rm LCDM} \dot \Delta = {\rho^{\rm LCDM} \Delta \over 2
m^2} \, .
\end{equation}
Thus, ordinary matter in mimicry models gravitates in agreement with the
effective value of the gravitational constant which appears in the cosmological
equations (\ref{past}) and (\ref{future}).

\begin{figure}
\begin{center}
\centerline{\psfig{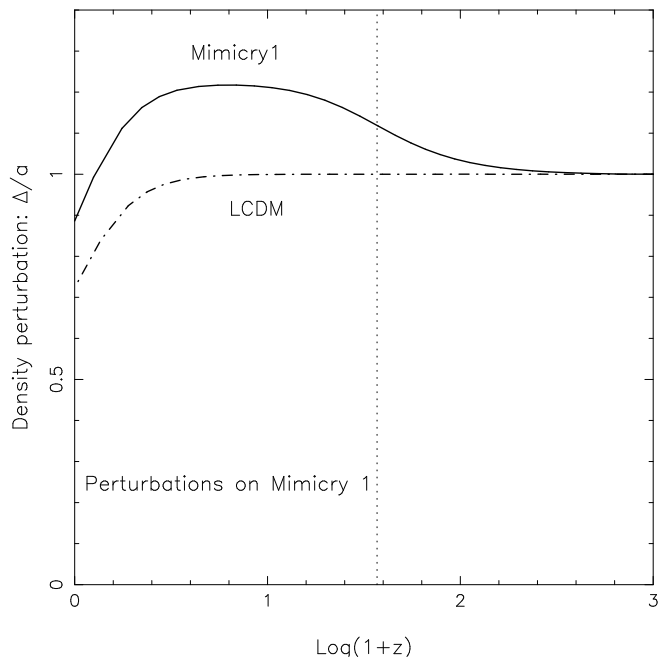} }
\end{center}
\caption{\small Density perturbations on the Mimicry\,1 brane (dot-dash) and in
LCDM (solid). The evolution of perturbations in Mimicry\,1 in this case is
effectively described by (\ref{matter}). In both cases, the perturbation
amplitude is scaled by the expansion factor $a(t)$. (It may be noted that
$\Delta/a = 1$ in standard CDM with $\Omega_\m = 1$.) The dotted vertical line
shows the mimicry redshift $z_\m \approx 37$. The braneworld has $\Omega_\m =
0.27$ while $\Omega^{\rm LCDM}_\m = 0.22$. This leads to a moderate enhancement
in the amplitude of brane perturbations over LCDM. \label{fig:growthB1}}
\end{figure}

We therefore conclude that, deep in the mimicry regime ($z \ll z_\m$),
perturbations grow {\em at the same rate\/} on the brane and in LCDM. This is
borne out by figure~\ref{fig:growthB1}, which shows the results of a numerical
integration of (\ref{matter}) for Mimicry\,1 [integrating the exact system
(\ref{one})--(\ref{three}) gives indistinguishable results]. Notice that the
{\em total\/} amplitude of fluctuations during mimicry in this model is {\em
greater on the brane\/} than in LCDM. Indeed, for mimicry models, we have
\begin{equation}
\frac{\Delta_{\rm brane}}{\Delta_{\rm LCDM}} \simeq
\frac{\Omega_\m}{\Omega^{\rm LCDM}_\m}  \quad \mbox{for} \quad z \ll z_\m \, ,
\label{ratio_omega}
\end{equation}
and this ratio is greater than unity for Mimicry\,1.

Since the contribution from perturbations in dark radiation can be neglected,
the growth of matter perturbations in Mimicry\,2 is again described by
(\ref{matter}) and by (\ref{ratio_omega}). However, since $\Omega_{\rm m} <
\Omega^{\rm LCDM}_\m$ in this case, the final amplitude of perturbations will
be {\em smaller\/} in Mimicry\,2 than the corresponding quantity in LCDM, which
is the opposite of what we have for Mimicry\,1.

It is interesting that during mimicry, when $\gamma \approx 1/3$, the relation
between the gravitational potentials $\Phi$ and $\Psi$ reduces to the
general-relativistic form $\Phi = \Psi$, as can be seen from
(\ref{difference}), where $\Phi$ satisfies the generalized Poisson equation
(\ref{potentials}), namely,
\begin{equation}
{1 \over a^2} \nabla^2 \Phi = \Theta {\Delta_\m \over 2 m^2} \, .
\end{equation}

An interesting feature of Mimicry\,1 is that, at early times, the expansion
rate in this model {\em exceeds\/} that in LCDM, i.e., $H(z)\vert_{\rm Mimicry
1} > H(z)\vert_{\rm LCDM}$ for $z > z_\m$ (see Figs.~\ref{fig:mimicry} \&
\ref{fig:theta}). [The opposite is the case for Mimicry\,2: the expansion rate
in this model is {\em lower\/} than that in LCDM at early times, i.e.,
$H(z)\vert_{\rm Mimicry\,2} < H(z)\vert_{\rm LCDM}$ for $z > z_\m$.]  As we can
see, this has important consequences for the growth of structure in this model.
The increase in the growth of perturbations in Mimicry\,1 relative to LCDM
occurs during the period before and slightly after the mimicry redshift has
been reached, when the relative expansion rate $H(z)/H^{\rm LCDM}$ is declining
while the `gravity term' $\Theta(z)$ has still not reached its asymptotic form
(\ref{theta1}). A lower value of $H(z)$ in (\ref{matter}) diminishes the
damping of perturbations due to cosmological expansion while a slower drop in
$\Theta(z)$ signifies a much more gradual decrease in the force of gravity.
Consequently, there is a net increase in the growth of perturbations on the
Mimicry\,1 brane relative to LCDM.\footnote{Figure~\ref{fig:theta} clearly
shows that $H(z)$ reaches its asymptotic form much sooner than $\Theta(z)$.
Notice that, at redshifts slightly larger than $z_\m$, the value of $\Theta(z)$
exceeds unity.  The dependence of perturbation growth on the mimicry redshift
$z_\m$ is very weak, and (\ref{ratio_omega}) is a robust result which holds to
an accuracy of better than $2\%$ for a wide range of parameter values.} For the
models in figure~\ref{fig:growthB1}, which have $\Omega^{\rm LCDM}_\m = 0.22$
and $\Omega_\m = 0.27$, the increase is about $20\%$. The increased amplitude
of perturbations in Mimicry\,1 stands in contrast to the DGP model as well as
Quintessence model, in both of which linearized perturbations grow at a {\em
slower\/} rate than in the LCDM cosmology \cite{LSS,KM,bernardeau}.

It is important to note that observations of galaxy clustering by the 2dFGRS
survey provide the following estimate \cite{2dF} for perturbation growth at a
redshift $z = 0.15$\,:
\begin{equation}
{d \log \delta \over d \log a} \equiv - \left. (1+z)\frac{d \log \delta}{dz}
\right\vert_{z = 0.15} = 0.51 \pm 0.11 \, , \label{eq:2dF}
\end{equation}
where $\delta \equiv \delta \rho / \rho$. Since the growth of perturbations
during the mimicry regime stays proportional to that in the LCDM model
($\delta_{\rm mimic} \propto \delta_{\rm LCDM}$, $z \ll z_\m$), it follows that
if perturbations in the LCDM model satisfy (\ref{eq:2dF}) (which they do), then so
will those in the mimicry scenario. Nevertheless, as we have seen, the {\em net
increase\/} in the amplitude of perturbations on the brane is {\em larger
than\/} that in the LCDM model. This clearly has important cosmological
consequences since it can enhance structure formation at high redshifts as well
as lead to higher values of $\sigma_8$. Thus, while preserving the many virtues
of the LCDM model, the mimicry models add important new features which could be
tested by current and future observations. We hope to return to some of these
issues in a future work.

\section{Discussion} \label{sec:discuss}

Braneworld theories with large extra dimensions, while having a number of very
attractive properties, also have a common difficulty:  on the one hand, the
dynamics of the higher-dimensional bulk space needs to be taken into account in
order to understand brane dynamics; on the other hand, all observables are
restricted to the four-dimensional brane.  In field-theoretic language, the
situation can be described in terms of an {\em infinite (quasi)-continuum\/} of
Kaluza--Klein gravitational modes existing on the brane from the brane
viewpoint.  This property makes braneworld theory complicated, solutions on the
brane non-unique and evolution nonlocal. For instance, while the solution with
a spherically symmetric source is relatively simple and straightforward in
general relativity, a similar problem in braneworld theory does not appear to
have a unique solution (see, e.g., \cite{MM}).

Fortunately, in situations possessing a high degree of symmetry, the above
properties of braneworld theory do not affect its cosmological solutions (at
least, in the simplest case of one extra dimension).  Thus, homogeneous and
isotropic cosmology on the brane is almost uniquely specified since it involves
only one additional integration constant which is associated with the mass of a
black hole in the five-dimensional bulk space. This makes braneworld theory an
interesting alternative for modelling dark-energy \cite{SS02} and dark-matter
\cite{SSV} effects on cosmological scales.

However, in order to
turn a braneworld model into a {\em complete\/} theory of gravity viable in
all physical circumstances, it is necessary to address the issue of boundary
conditions.  Usually, one tries to formulate reasonable conditions in the bulk
by demanding that the bulk metric be nonsingular or by employing other
(regulatory) branes.  However, neither of these conditions have been
implemented in braneworld theory in full generality; moreover, they leave open
the problem of nonlocality of the gravitational laws on the brane
(since the brane is left open to influences from the bulk).

In this paper, we adopted a different approach to the issue of boundary
conditions in the brane--bulk system.  From a broader perspective, boundary
conditions can be regarded as any conditions restricting the space of
solutions. Our approach is to specify such conditions directly on the brane
which represents the observable world, in order to arrive at a local and closed
system of equations on the brane. The behaviour of the metric in the bulk is of
no further concern in this approach, since this metric is for all practical
purposes unobservable. Since the nonlocality of the braneworld equations is
known to be connected with the dynamical properties of the bulk Weyl tensor
projected onto the brane \cite{SMS}, it is natural to consider the possibility
of imposing certain restrictions on this tensor. Perhaps, the simplest
condition is to set to zero its (appropriately defined) anisotropic stress.
This is fully compatible with all the equations of the theory and results in a
brane universe described by a modified theory of gravity and having an
additional invisible component --- dark radiation --- which is endowed with
nontrivial dynamics. More generally, we suggest (\ref{bc}) and
(\ref{bc-linear}) which describes a one-parameter family of BC's with the
parameter $\bc$. This family generalizes the boundary condition derived by
Koyama and Maartens \cite{KM} for the DGP model in the small-scale and
quasi-static approximation (Koyama and Maartens derived the value $\bc =
-1/2$).

An important conclusion of our paper is that the growth of perturbations in
braneworld models strongly depends upon our choice of BC's.  This was
illustrated in figure~\ref{fig:bc} for the DGP model. Specifying boundary
conditions in the form (\ref{bc-linear}) allows us to determine regions of
stability and instability in terms of the single parameter $\bc$; they are
described by Eq.~(\ref{stable}). In the DGP model, perturbations are explicitly
demonstrated to be quasi-stable for $\bc = 1/2$ (figure~\ref{fig:dgp-stable})
and unstable for $\bc = -1/2$ (figure~\ref{fig:dgp}). In the instability
domain, gradients in the momentum potential $v_\C$ of dark radiation, lead to
the creation of perturbations in this quantity via equation (\ref{two}). This
effect can significantly boost the growth of perturbations in matter. An
important implication of this effect is that perturbations in the baryonic
component might overcome the `growth problem,' which plagues them in standard
general relativity, and grow to acceptable values without requiring the
presence of (deep potential wells in) dark matter. The Mimicry\,2 model looks
promising from this perspective. Note that, in this model, the expansion of a
low-density universe is virtually indistinguishable from that of a
(higher-density) LCDM model. Enhanced perturbation growth in this `mimicry'
model might permit a low-density universe to also match observations of
gravitational clustering (thereby circumventing the need for large amounts of
dark matter to achieve this purpose).\footnote{Of course, a valid model of
structure formation should also agree with observations of gravitational
lensing which appear to require some amount of dark matter \cite{bullet}, as
well as provide a good fit to the observed power spectrum of galaxies and
fluctuations in the cosmic microwave background; see \cite{loiter1} for
discussions of this issue within a different context.} A detailed investigation
of this effect, however, lies outside the scope of the present paper and will
be taken up elsewhere.

Values of $\bc$ lying in the stability region (\ref{stable}) or a scale-free
boundary condition such as (\ref{bc-new}) may also be important. In this case,
perturbations in the DGP model grow {\em slower\/} than in LCDM whereas
perturbations in Mimicry\,1 grow somewhat faster. This suggests that structure
formation may occur slightly earlier in Mimicry\,1 than it does in LCDM.

Two related points deserve mention. In general relativity, the linearized
perturbation equation
\begin{equation}
{\ddot \delta} + 2H {\dot\delta} = \frac{3}{2} H_0^2 \Omega_\m \left(
\frac{a_0}{a} \right )^3 \delta \label{perturb_last}
\end{equation}
depends exclusively upon the expansion history of the universe and the value of
$\Omega_\m$, so that a knowledge of $H (z)$ from independent sources (such as
the luminosity distance) permits one to determine $\delta (z)$ and, vice-versa,
the observed value of $\delta (z)$ can be used to reconstruct the expansion
history $H (z)$ \cite{ss06}.  As we have demonstrated in this paper, equation
(\ref{perturb_last}) is no longer valid for the braneworld, and it remains to
be seen whether knowing $H(z)$ and $\Omega_\m$ will allow us to reconstruct
$\delta (z)$ uniquely.

It is also well known that the expansion history, $H(z)$, does not characterize
a given world model uniquely, and it is conceivable that cosmological models
having fundamentally different theoretical underpinnings (such as different
forms of the matter Lagrangian or different field equations for gravity) could
have identical expansion histories (some examples may be found in \cite{ss06}
which also contains references to earlier work). As an illustration, consider
the Brane\,2 model in which the effective equation of state is never
phantom-like, so that $w_{\rm eff} \geq -1$. As we have seen, Mimicry\,2 and
the DGP model form important subclasses of this braneworld. A quintessence
potential corresponding to a given expansion history can be determined from
\cite{star}
\begin{eqnarray}
{8\pi G\over 3H_0^2} V(x)\ &=& {H^2 \over H_0^2} - {x\over 6H_0^2} {dH^2 \over
dx} - \frac12 \Omega_\m\,x^3 \, , \label{eqn:Vzed}\\
{8\pi G \over 3H_0^2} \left( {d\phi \over dx} \right)^2 &=& {2 \over 3H_0^2 x}
{d\ln H \over dx} - {\Omega_\m x \over H^2} \, , \quad x \equiv 1+z \, .
\label{eqn:phidot}
\end{eqnarray}
Substitution for $H(z)$ from (\ref{rw}) with the plus sign results in a
quintessence model which has {\em the same expansion history\/} as the
braneworld. Perturbative effects, however, are likely to break this degeneracy,
since it is unlikely that matter perturbations will grow at the same rate in
the quintessence model as on the brane.

Although we have explicitly solved the perturbation equations only for two
cases: the mimicry models \cite{SSV} and the DGP braneworld \cite{DGP}, it is
quite clear that the treatment developed by us may have larger ramifications
since it could be applied to a broader class of models. In \cite{timelike}, for
instance, it was shown that a braneworld dual to the Randall--Sundrum\,II model
--- the bulk dimension being time-like instead of space-like --- will bounce at
early times, thereby avoiding the singularity problem which plagues general
relativity. It would be an interesting exercise to apply the formalism
developed in this paper to this non-singular braneworld to see if the problem
of tachyonic gravitational modes can be avoided. Since the boundary conditions
of type (\ref{weyl})--(\ref{bc}), specified directly on the brane, effectively
freeze certain degrees of freedom in the brane--bulk system, the corresponding
braneworld theory may also be free from the debatable problem of ghosts (see
\cite{ghost} for a recent discussion of this problem).

Whether boundary conditions such as those described by equations
(\ref{weyl})--(\ref{bc}) or (\ref{bc-new}) will remain in place for a more
fundamental extra-dimensional theory is presently a moot point. Perhaps, by
comparing the consequences of different boundary conditions with observations
we  will gain a better understanding of the type of braneworld theory most
consistent with reality.

We believe that our calculations have opened up an interesting avenue for
further research in braneworld cosmology. Related astrophysical phenomena which
demand further exploration include gravitational lensing, fluctuations in the
cosmic microwave background, the power spectrum of density fluctuations, etc.
For a deeper understanding of gravitational instability on the brane, we also
need to examine the behaviour of perturbations during the radiative epoch as
well as their origin as vacuum fluctuations (presumably) during inflation.
Additional issues of considerable interest relate to non-linear gravitational
clustering on the brane. For instance, it is well known that a density field
which was originally Gaussian develops non-Gaussian features even in the weakly
non-linear regime of gravitational instability \cite{non-gauss}. Although these
effects do not appear to be unduly sensitive to the presence of dark energy,
this has only been tested for $\Lambda$CDM and a few other conventional
dark-energy models \cite{tatekawa}. The value of the lower moments of the
density field such as its skewness
$\langle\delta^3\rangle/\langle\delta^2\rangle^2$ and kurtosis
$\langle\delta^4\rangle/\langle\delta^2\rangle^3$ in models which give rise to
acceleration because of amendments to the {\em gravitational\/} sector of the
theory remain unexplored and could lead to a deeper understanding of structure
formation in these models.

We hope to return to some of these issues in the future.

\section*{Acknowledgments}

The authors acknowledge useful correspondence with Roy Maartens and an
insightful conversation with Tarun Deep Saini. This work was supported by the
Indo-Ukrainian program of cooperation in science and technology sponsored by
the Department of Science and Technology of India and Ministry of Education and
Science of Ukraine.  Yu S and A V also acknowledge support from the Program of
Fundamental Research of the Physics and Astronomy Division of the National
Academy of Sciences of Ukraine and from the INTAS grant No.~05-1000008-7865.

\appendix
\section{Attempts at deriving boundary conditions}

Recent attempts at deriving proper boundary conditions for the brane--bulk
system by considering perturbations in the bulk were made in \cite{KM,SSH}.
Here we would like to discuss this approach in more detail and illustrate some
difficulties with this treatment.

Both approaches study perturbation in the DGP model, which has Minkowski bulk
background, and use the Mukohyama master equation \cite{Mukohyama} for the
scalar variable $\Omega (t, {\vec x}, y)$ written in Gaussian normal
coordinates with respect to the brane, which is located at $y = 0$ in the fifth
coordinate $y$ and has intrinsic coordinates $\{t, {\vec x} \}$:
\begin{equation} \label{Mukohyama}
- {\partial \over \partial t} \left( {1 \over n b^3} {\partial \Omega \over
\partial t} \right) + {\partial \over \partial y} \left( {n \over b^3} {\partial \Omega \over
\partial y} \right) - {n k^2 \over b^5} \Omega = 0 \, ,
\end{equation}
where
\begin{equation}
n (t, y) = 1 + \left( {\dot H \over H} + H \right) y \, , \qquad b(t, y) = a
\left( 1 + H y \right)
\end{equation}
are the functions describing the unperturbed bulk metric
\begin{equation}
ds^2 = - n^2 dt^2 + b^2 d{\vec x}^2 + dy^2 \, ,
\end{equation}
and $k$ is the wave number in the Fourier decomposition in the ${\vec x}$
space.

The (suitably normalized) Mukohyama variable is related to perturbations on the
DGP brane via \cite{Deffayet}:
\begin{equation} \label{relations}
\begin{array}{l}
\displaystyle \delta \rho_\C  = -  { k^4 \over 3 a^5} \left. \Omega
\right|_{y = 0} \, , \medskip \\
\displaystyle v_\C  = { k^2 \over 3 a^3 } \left. \left( \dot
\Omega - H \Omega \right) \right|_{y = 0} \, , \medskip \\
\displaystyle \delta \pi_\C  = - { 1 \over 6 a} \left. \left( 3 \ddot \Omega -
3 H \dot \Omega + { k^2 \over a^2} \Omega + 3 \dot H \left( \ell - H^{-1}
\right) {\partial \Omega \over \partial y} \right) \right|_{y = 0} \, .
\end{array}
\end{equation}

Koyama and Maartens \cite{KM} study perturbations on subhorizon scales, $k \gg
aH$, and employ what they call `quasi-static approximation,' which consists in
neglecting the time derivatives in (\ref{Mukohyama}) and (\ref{relations}). By
considering the approximate `quasi-static' solution for $\Omega$ with
appropriate (decaying) boundary condition in the bulk, they argue that the term
with the derivative $\partial \Omega / \partial y$ in the brackets on the
right-hand side of the last relation in (\ref{relations}) can also be
neglected, and this relation then becomes
\begin{equation} \label{KM-relations}
\delta \pi_\C  \approx - {k^2 \over 6 a^3} \left. \Omega \right|_{y = 0} \, .
\end{equation}
This results in the Koyama--Maartens boundary condition (\ref{bc-linear}),
(\ref{KM}).

However, as was shown in the paper, the boundary condition (\ref{bc-linear}),
(\ref{KM}) leads directly to equation (\ref{closed}), which exhibits strong
scale-dependent instability. The last effect was also demonstrated numerically.
We note that, in deriving equation (\ref{closed}), we used the exact system of
equations (\ref{one})--(\ref{three}) on the brane, and we did not perform any
time differentiation of the approximate relation (\ref{bc-linear}), which might
be illegitimate in the quasi-static approximation. This means, in our view,
that the quasi-static approximation, which was {\em assumed\/} in deriving
conditions (\ref{KM-relations}), {\em cannot hold\/} during the whole course of
evolution described by the system of equations for perturbations
(\ref{one})--(\ref{three}).

Sawicki, Song, and Hu \cite{SSH} study scalar perturbations on all scales. They
note that the distance $y_{\rm hor}$ to the causal horizon of the brane remains
proportional to $H^{-1}$ during periods of domination of matter with constant
equation of state, and they impose the zero boundary condition for $\Omega$ at
this causal horizon.  This does not fix the value of $\Omega$ at the brane, and
it is done separately by setting a `scaling' ansatz of the form
\begin{equation} \label{ansatz}
\Omega  = A(p) a^p G \left(x, {k \over aH} \right)\, , \qquad x = {y \over
y_{\rm hor}} \, , \qquad y_{\rm hor} = \xi H^{-1} \, .
\end{equation}
They substitute (\ref{ansatz}) into (\ref{Mukohyama}) to derive a differential
equation on $G \left(x, {k \over aH} \right)$ during periods of domination of a
fluid with constant equation of state (radiation or matter). In doing this,
they treat $p$, $A(p)$, and $\xi$ as constants and also neglect the derivative
of $G \left(x, {k \over aH} \right)$ with respect to the second argument. One
cane note that this already imposes certain restriction on the form of $\Omega$
on the brane, hence, via (\ref{relations}), also on the perturbations of dark
radiation. The result is the differential equation for $G \left(x, {k \over aH}
\right)$ of the form
\begin{equation} \label{g}
{\partial^2 G \over \partial x^2} + f_1 (p, x) {\partial G \over \partial x} +
\left[ f_0 (p, x) - \left({k \over aH} \right)^2 g(p, x) \right] G = 0 \, ,
\end{equation}
where $f_0$, $f_1$, and $g$ are some functions, and in which the quantity ${k
\over aH}$ plays the role of a parameter.  This equation is solved with the
boundary conditions $G(0) = 1$, $G(1) = 0$, which uniquely determines the
derivative $G'(0)$, hence, the quantity $\partial \Omega / \partial y$ on the
brane needed in (\ref{relations}).

The power $p$ in (\ref{ansatz}) is not yet specified, and it is adjusted in
\cite{SSH} by an iterative procedure so as to satisfy the Bianchi identity on
the brane coupled to the cosmological equations for matter perturbations. In
this way, $p$ also becomes a function of the scale factor $a$.

One can see that a number of assumptions were made in this procedure. In
deriving equation (\ref{g}), the time derivatives of $p$ and the partial
derivative of $G \left(x, {k \over aH} \right)$ with respect to its second
argument were neglected.  Together with ansatz (\ref{ansatz}), this means that
an effective `quasi-static' approximation was used for $\Omega$, similar to the
one employed in \cite{KM}.  It is not surprising, therefore, that, on
subhorizon scales, the result of \cite{KM} is reproduced, with the approximate
relations (\ref{bc-linear}), (\ref{KM}), for which the quasi-static regime is
unstable, as we have shown in this paper.

Furthermore, the authors did not extend smoothly their approximate solution in
the bulk beyond the horizon $y = y_{\rm hor}$, so it is not at all clear
whether it is really singularity-free in the bulk, as was claimed.  To specify
simply $\Omega \equiv 0$ for $y \ge y_{\rm hor}$ means introducing a
discontinuity in the derivative of $\Omega$ at $y = y_{\rm hor}$, which itself
can be regarded as a singularity.  This once again illustrates the general
difficulty which one encounters in finding singularity-free solutions in the
bulk.

\end{document}